\documentclass[lettersize,journal]{IEEEtran}
\usepackage{amsmath,amsfonts}
\usepackage{algorithmic}
\usepackage{array}
\usepackage[caption=false,font=normalsize,labelfont=sf,textfont=sf]{subfig}
\usepackage{textcomp}
\usepackage{stfloats}
\usepackage{url}
\usepackage{verbatim}
\usepackage{graphicx}
\usepackage{bm}
\usepackage{color}
\usepackage{graphicx}
\usepackage{lipsum}
\usepackage{amsmath}
\usepackage{amssymb}
\usepackage{mathrsfs}
\usepackage{algorithm}
\usepackage{array}
\usepackage{fixltx2e}
\usepackage{stfloats}
\usepackage{url}
\usepackage{booktabs}
\usepackage{enumerate}
\usepackage{setspace}
\usepackage{float}
\def\BibTeX{{\rm B\kern-.05em{\sc i\kern-.025em b}\kern-.08em
    T\kern-.1667em\lower.7ex\hbox{E}\kern-.125emX}}
\usepackage{balance}
\usepackage{amsthm}
\usepackage{hyperref}
\usepackage{makecell}

\definecolor{CBLUE}{RGB}{0,112,192}
\newtheorem{theorem}{Theorem}[section]
\newtheorem{lemma}[theorem]{Lemma}
\newtheorem{definition}[theorem]{Definition}

\newtheorem{proposition}[theorem]{Proposition}

\newtheorem{remark}[]{Remark}

\theoremstyle{definition} 
\newtheorem{example}{Example}
\begin{document}
\title{Quantifying Grid-Forming Behavior: Bridging Device-Level Dynamics and System-Level Strength}
\author{Kehao Zhuang, Huanhai Xin, {\em Senior Member, IEEE}, Verena Häberle, {\em Member, IEEE}, Liangxiao Luo, Xiaoying Liu, Xiuqiang He, {\em Member, IEEE}, Linbin Huang, {\em Member, IEEE}, and Florian Dörfler, {\em Fellow, IEEE}
\vspace{-8mm}

\thanks{This work was supported by the Science and Technology Programme of the State Grid Corporation under Grant 5226SX25001R-384-ZN.}
\thanks{H. Xin, L. Luo, L. Huang, and X. Liu are with the College of Electrical Engineering, Zhejiang University, Hangzhou, China. (Emails: \{xinhh, luolx, hlinbin\}@zju.edu.cn, liu\_xiaoying2023@163.com).}
\thanks{K. Zhuang, V. Häberle, F. Dörfler, are with the Automatic Control Laboratory at ETH Zürich, Switzerland. (Emails: \{kzhuang, verenhae, dorfler\}@ethz.ch).}
\thanks{X. He is with the Department of Automation, Beijing National Research Center for Information Science and Technology, Tsinghua University, Beijing, China (e-mail: xhe@tsinghua.edu.cn).}
}

\maketitle

\begin{abstract}
Grid-forming (GFM) technology is widely regarded as a promising solution for future power systems dominated by power electronics. However, {a universally accepted definition of GFM behavior and precise method for its quantification remain elusive.} Moreover, the impact of {GFM converters} on system stability is not precisely quantified, creating a significant disconnect between device and system levels. To address these gaps from a  small-signal perspective, at the device level, the paper introduces a novel metric, the \textit{Forming Index} ($FI$) to quantify a converter's response to grid voltage fluctuations. Rather than enumerating various control architectures, the $FI$ provides a metric for the converter's GFM ability by quantifying its sensitivity to grid variations. At the system level, a new quantitative measure of {\em system strength} that captures the multi-bus voltage stiffness is proposed, which quantifies the voltage and phase angle responses of multiple buses to current or power disturbances. The paper further extends and defines this concept to {\em grid strength} and {\em bus strength} to identify weak areas within the system.  Finally, the device and system levels are bridged by formally proving that GFM converters enhance system strength. The proposed framework provides a unified benchmark for GFM converter design, optimal placement, and system stability assessment.
\end{abstract}


\vspace{-4mm}
\section{Introduction}
\IEEEPARstart{S}{table} voltage and frequency are basic requirements for reliable power system operation. In conventional power systems, synchronous generators (SGs) naturally act as voltage sources, providing inherent voltage and frequency support~\cite{kunder:stability_classification}. However, the increasing integration of renewable energy sources through power electronic converters is gradually reducing the presence of SGs, thereby diminishing their contribution to system stability. Unlike SGs, converters are typically controlled by constant power control and phase-locked-loop (PLL)~\cite{xiongfei:syn_overview}. As PLL-based converters have gradually displaced SGs, weak grid characteristics with low short circuit ratio (SCR) and low inertia are more pronounced, posing new challenges to power system security~\cite{20:MilanoPSCC}. 

To address these challenges and improve power system stability, {converters have been designed to emulate the characteristics of synchronous generators (SGs), thereby actively establishing stable voltage and frequency. Typical examples include virtual synchronous generator (VSG) control and droop control, and such converters are commonly referred to as grid-forming (GFM) converters~\cite{xiongfei:syn_overview}. In addition to these SG-emulating approaches, other GFM controls have been proposed, such as dispatchable virtual oscillator control (dVOC)~\cite{3:XIUQIANGDVOC}. Within this development, a common practice has been to distinguish GFM and grid-following (GFL) converters based on their synchronization control structures, where PLL-based converters are typically classified as GFL converters~\cite{Thomas:GFM_GFL_comparison}.}

{However, this classification, although simple, is not sufficiently comprehensive, as it relies on control structure rather than actual dynamic behavior. For example, \cite{9:pll-gfm, PLLGFM} propose adding virtual admittance control to a PLL-based converter (referred to as PLL-VAC), which can achieve dynamic behavior similar to that of a VSG. In the original paper, this approach is termed PLL-based GFM control. This indicates that converters with different synchronization structures may exhibit similar dynamic behavior after appropriate modifications.}

Therefore, in order to further refine the concept of GFM converter without enumerating every control strategy, various power organizations have issued reports on GFM behaviors \cite{GFM:NERC,gfm:AEMO,gfm:ACER,gfm:unifi}. As summarized in Table~\ref{tab:GFM}, despite variations in the definitions, a widely recognized and crucial behavior of GFM converters is their ability to exhibit \textit{stiff voltage source behavior} across sub-transient to transient timescales. 

\begin{table}
    \centering
    \caption{GFM functionality of different reports}
    \begin{tabular}{|c|c|}\hline
        Report & GFM functionality\\\hline
         \makecell{2021,NERC\\USA\cite{GFM:NERC}} & \makecell{``maintaining an {\em internal voltage phasor} \\that is constant or nearly constant in\\ the  sub-transient to transient time frame"}\\\hline
        \makecell{2023,AEMO\\Australia\cite{gfm:AEMO}}& \makecell{``maintains a constant {\em internal voltage} \\ {\em phasor} in a short time frame, with \\ magnitude and frequency ... immediate \\ response to a change in the {\em external grid}"} \\\hline
        \makecell{2023,ACER\\Europe\cite{gfm:ACER}}& \makecell{``behaving at the terminals of the individual \\ unit(s) as a {\em voltage source behind an} \\ {\em internal impedance} (Thevenin source)''} \\\hline
        \makecell{2024,UNIFI\\USA\cite{gfm:unifi}} & \makecell{``maintaining an {\em internal voltage phasor}\\ that is constant or nearly constant in the \\ sub-transient to transient time frame"} \\ \hline
    \end{tabular}
    \label{tab:GFM}
    \vspace{-3mm}
\end{table}

{Although GFM behavior is commonly defined as stiff voltage source behavior, such a definition remains qualitative, and existing approaches for quantifying this behavior are still not sufficiently developed.} References~\cite{6:Debryfrequencysmooth} and~\cite{21:grossfrequencysmooth} propose the concept of \textit{frequency smoothing} capability as a necessary condition for {GFM behavior}, which measures the sensitivity of a converter's frequency to grid frequency variations, {i.e., its ability to follow or reject grid frequency disturbances~\cite{21:grossfrequencysmooth}.} However, frequency smoothing only quantifies the rigidity of the converter's frequency.
To address both voltage and frequency characteristics, reference \cite{7:howmany} introduced the maximum singular values of the converter impedance matrix. It demonstrated that a VSG and a PLL-based converter with AC voltage control (PLL-PV) act as two-dimensional (stiff voltage and frequency) and one-dimensional (only stiff voltage) voltage sources, respectively. In addition, voltage source characteristics can also be evaluated by assessing the impedance matching between the converter and an ideal line with resistance and inductance \cite{8:eduardovoltage}. While these studies attempt to quantify GFM behavior by measuring how closely a converter approximates an ideal voltage source, a clear boundary between GFM and GFL behaviors remains undefined. Specifically, it is unclear what degree of approximation to a stiff voltage source qualifies a converter as a GFM converter. 

Moreover, another often ducked question is why power systems require GFM converters. A common perception is that power systems need inertia and synchronization \cite{GFM:keytechonology,kehao:dual_axis}, yet both GFM and GFL converters can provide that \cite{1:adaptive}. An alternative perspective suggests that GFM converters have the potential to enhance \textit{system strength} \cite{Huanhai:place_GFM}. However, due to the absence of clearly defined quantitative metrics for both GFM behavior and system strength, this hypothesis has lacked a formal proof. Therefore, from the perspective of ensuring stable power system operation, the fundamental difference between GFM and GFL behaviors is insufficiently explored.

In order to fill the above research gap, this paper quantitatively answers three key questions from a small-signal perspective: what is a {GFM converter}, why is it needed, and how does it relate to system strength? The main contributions are as follows:

1) At the device level, {the {\em Forming Index (FI)} is proposed} based on the concept of frequency smoothing and defined as the maximum singular value of the sensitivity function from the grid voltage to the converter voltage. This index quantifies the extent to which the converter either follows or rejects grid variations, thereby reflecting its voltage source behavior. {This provides a clear distinction between GFM and GFL behaviors, thereby enabling accurate classification of control strategies (or converters) that exhibit GFM behavior as GFM control (or GFM converters) from the small-signal perspective.} Different control architectures are validated using $FI$, demonstrating that GFM behavior can also be realized based on PLL.

2) At the system level, { the concept of {\em system strength} is formally introduced} as a quantitative criterion for the requirements of stable power system operation. This metric is defined as the sensitivity of the multi-bus voltage vector to multi-bus current (or power) disturbances, thereby capturing the overall voltage stiffness of the system. The definitions for {\em grid strength} and {\em bus strength} are also provided to accurately identify weak buses, which generalize the well known SCR. 

3) In a small-signal setting, a formal theoretical proof establishes the link between the device and system levels: namely, a converter exhibiting GFM behavior enhances system strength.  Furthermore, this paper also discusses how the $FI$ and system strength can be employed to formulate GFM control design and placement as optimization models.

{\em Notation}: Let $\mathbb{R}$ and $\mathbb{C}$ denote the set of real numbers and the set of complex numbers respectively; $\mathcal{RH}_{\infty}$ denotes real-rational, stable, and proper function space with bounded $\mathcal{H}_\infty$ norm; $\mathfrak{R}(\cdot)$ denotes take the real part; $\Delta$ denotes the small-signal perturbation increment; $\sigma(\cdot)$ and $\lambda(\cdot)$ denote the singular value and eigenvalue respectively;  $\overline{(\cdot)}$  and $\underline{(\cdot)}$denote the maximum and minimum values, respectively. For a matrix $A\in \mathbb{C}^{n\times n}$, let $A^H$ denote the conjugate transpose of $A$, {let ${\rm Re}(A)=\underline{\lambda}\left(\frac{A+A^H}{2}\right)$.} Let ${\rm det}(\cdot)$ denote the determinant of a matrix.

{\em Inequalities}:  $A_{ij}$ is a block of $A$, and it has $\overline{\sigma}(A)\leq {\rm max}_{\forall i} \sqrt{\textstyle \sum_{j=1}^{n}\overline{\sigma}(A_{ij})\textstyle \sum_{j=1}^{n}\overline{\sigma}(A_{ji})}\leq{\rm max}_{\forall i}\left[\textstyle \sum_{j=1}^{n}\overline{\sigma}(A_{ij})\textstyle \sum_{j=1}^{n}\overline{\sigma}(A_{ji})\right]$, $\underline{\sigma}(A)\geq \underline{\lambda}(\frac{A+A^H}{2})$. For a matrix $B\in \mathbb{C}^{n\times n}$, there are $\underline{\sigma}(A+B)\geq\underline{\sigma}(A)-\overline{\sigma}(B)$, $\underline{\sigma}(AB)\geq\underline{\sigma}(A)\underline{\sigma}(B)$ and $\overline{\sigma}(AB)\leq\overline{\sigma}(A)\overline{\sigma}(B)$.  

\section{{Device Level: GFM Behavior of Converters}} 
\vspace{0mm}

\begin{figure}
	\centering
	\includegraphics[width=3.3in]{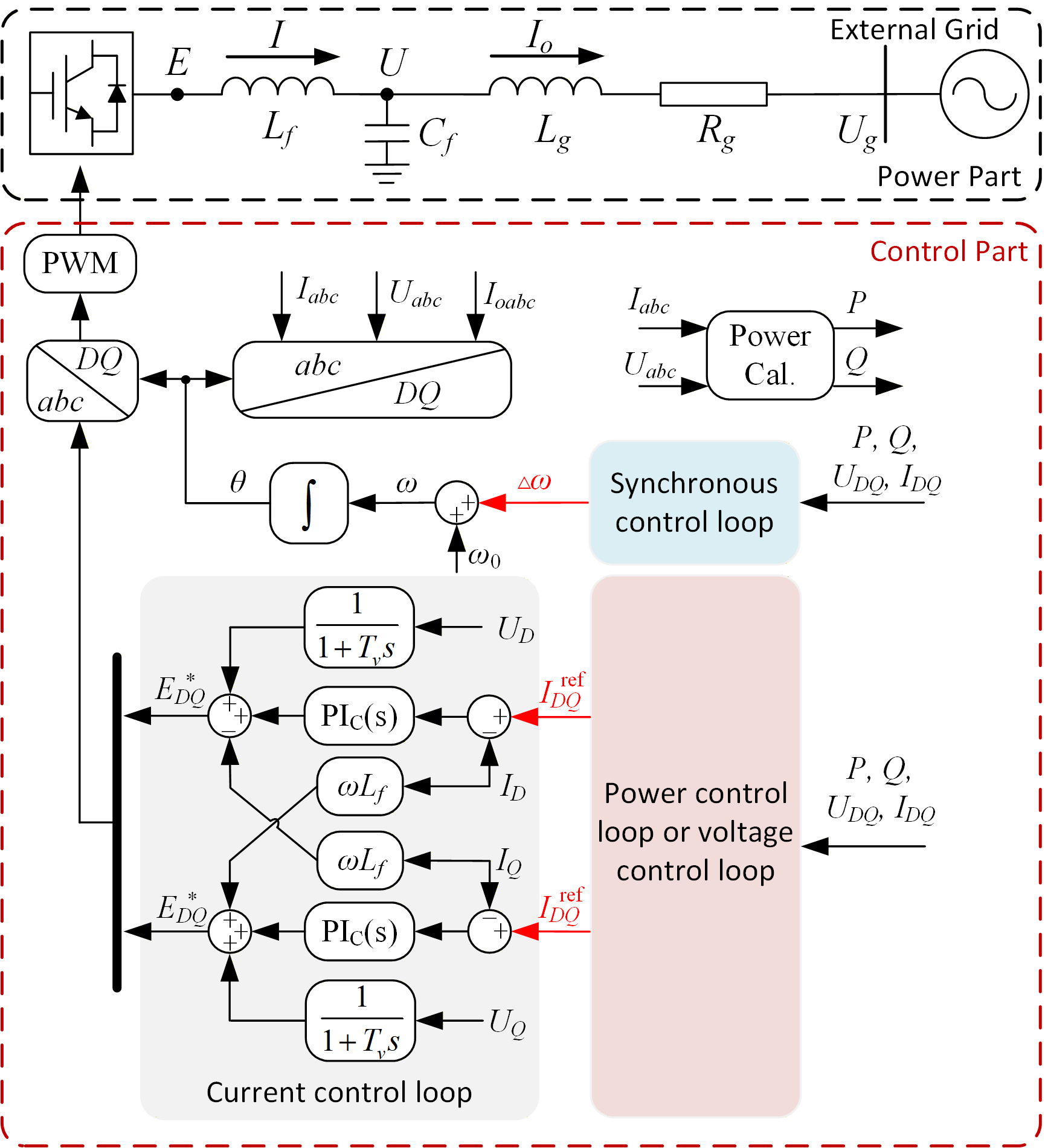}
	\vspace{-3mm}
	\caption{The diagram of a single converter connected to the grid.} 
	\vspace{-0.2cm}
	\label{fig1:converter}
\end{figure}

\begin{figure} 
	\centering
	\includegraphics[width=3.5in]{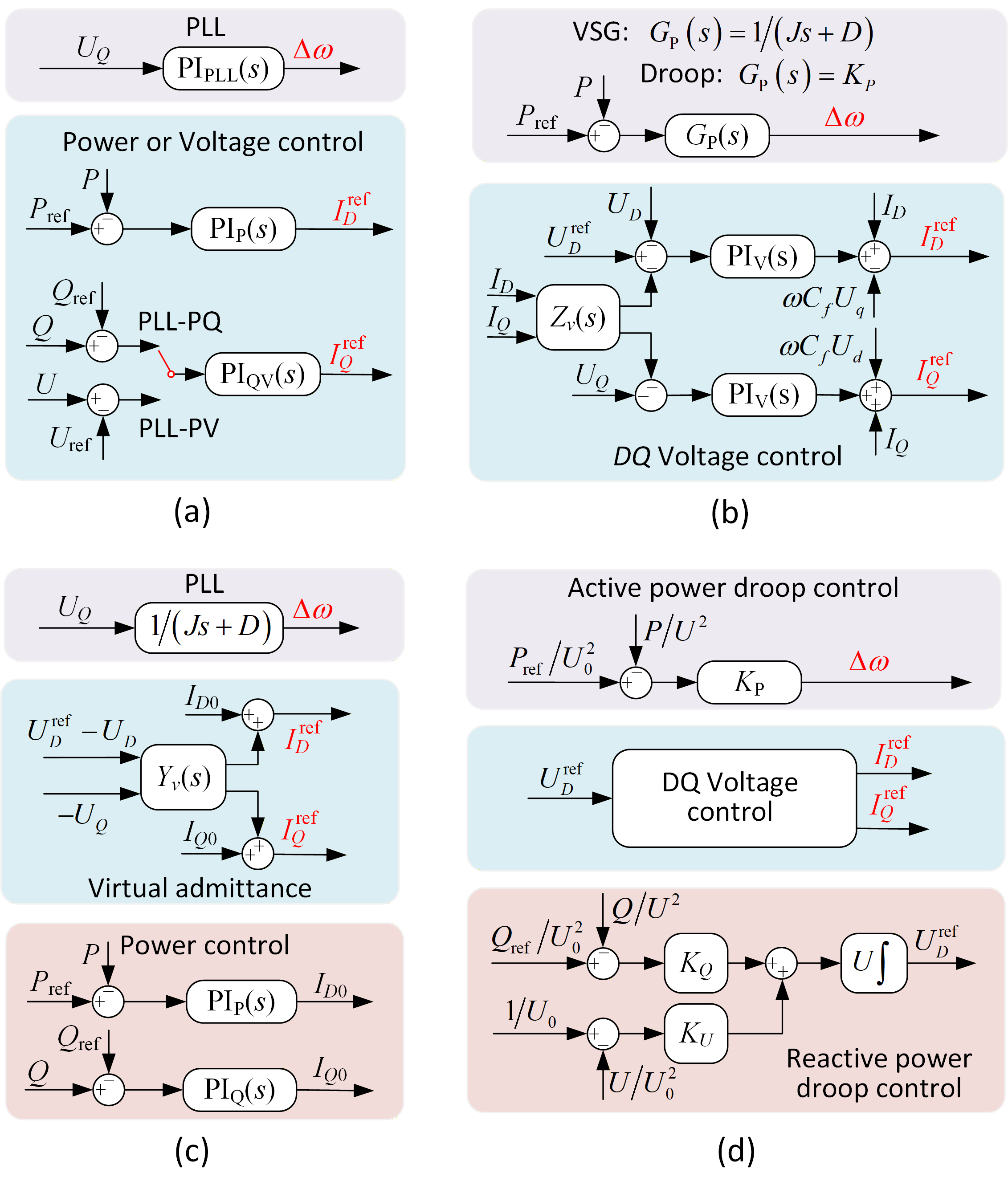}
	\vspace{-3mm}
	\caption{The common control strategies. (a) PLL-based converter with reactive control (PLL-PQ) or voltage control (PLL-PV). (b) VSG or droop control (droop). (c) {PLL-based converter with virtual admittance control (PLL-VAC).} (d) dVOC.} 
		\vspace{-4mm}
	\label{fig2:control}
\end{figure}

At the device level, as shown in Fig.~\ref{fig1:converter}, consider a single converter connected to the external grid modeled as a stiff voltage source. The converter output voltage, grid voltage, and output current vector in the global dq frame with a constant nominal rotational frequency are denoted as $U_{dq}$, $U_{{\rm g}dq}$ and $I_{dq}$, respectively. The subscripts $_{D}$ and $_{Q}$ in Figs.~\ref{fig1:converter} and \ref{fig2:control} indicate variables expressed in the converter’s local $dq$ frame.  The inductance and resistance of the interconnecting line are $L_{\rm g}$ and $R_{\rm g}$, respectively. 
 
	\vspace{-2mm}
\subsection{Linearized model of a converter connected to the grid}

After linearization, the small-signal dynamics of a converter with any of the control architectures shown in Fig.~\ref{fig2:control} can be represented by its admittance transfer function matrix \cite{guyunjie:impedance}. This matrix describes the relationship between the converter's terminal voltage vector $\Delta{ U}_{{dq}}=\begin{bmatrix} \Delta U_d & \Delta U_q \end{bmatrix}^{\top}$ and the current vector $\Delta{ I}_{{dq}}=\begin{bmatrix} \Delta I_d & \Delta I_q \end{bmatrix}^{\top}$ (the direction of current flowing into the converter is defined as positive, and is normalized with respect to the converter’s rated capacity):

\vspace{-1mm}
{\begin{equation}\label{eq1:Yvsc}
\Delta{I}_{{dq}}= 
e^{J\delta}{Y}_{\rm L}(s)e^{-J\delta}\Delta{U}_{{dq}}=:
{Y}_{\rm de}(s)\Delta{U}_{{dq}} \,,
\end{equation}
where ${Y}_{\rm L}(s)$ is a $2\times 2$ transfer function matrix describing the converter’s dynamics using local per-unit calculations, ${Y}_{\rm de}(s)$ is a $2\times 2$ admittance function matrix in global dq frame. $\delta$ is the steady-state angle difference between the global dq frame and the converter’s local dq frame, and $e^{J\delta}=\left[\begin{smallmatrix} {\rm cos}\delta & -{\rm sin}\delta \\ {{\rm sin}\delta} & {\rm cos}\delta \end{smallmatrix}\right]$.  

{ The converter admittance models ${Y}_{\rm de}(s)$ used in the subsequent analysis are all obtained by linearizing the complete dynamic models shown in Figs. 1 and 2, as detailed in Appendix~\ref{sec:model}. }

The dynamics of the line are
\vspace{-1mm}
\begin{equation}\label{eq2:Yline}
\Delta{I}_{{dq}}= \frac{1}{L_{\rm g}}
\underbrace{\begin{bmatrix}
  \frac{s}{\omega_0}+\tau  & -1 \\ 1 & \frac{s}{\omega_0}+\tau
\end{bmatrix}^{-1}}_{=:{\gamma}(s)}
\left(\Delta{U}_{{\rm g}dq}-\Delta{ U}_{{dq}}\right) \,,
\end{equation}
where $\tau=R_{\rm g}/L_{\rm g}$ is the ratio of line resistance to inductance,  $\Delta { U}_{{\rm g}dq}=\begin{bmatrix} \Delta U_{{\rm g}d}& \Delta U_{{\rm g}q}\end{bmatrix}^{\top}$, ${\gamma}(s)$ is the line dynamic matrix.

By combining~\eqref{eq1:Yvsc} and \eqref{eq2:Yline}, the closed-loop dynamics are modeled as the feedback interconnection shown in Fig. \ref{fig3:singleclose}. The transfer function ${S}_v(s)$ from the grid voltage disturbance vector $\Delta U_{{\rm g}dq}$ to the converter voltage vector $\Delta { U}_{dq}$ termed the {\em sensitivity function}, is
\begin{equation}\label{eq3:sensitivity of single VSC}
\Delta{U}_{dq}=\underbrace{\left[{I}_2+L_{\rm g}{\gamma}^{-1}(s){Y}_{\rm de}(s)\right]^{-1}}_{=:{S} _v(s)}\Delta{ U}_{{\rm g}dq}\,,
\end{equation} 
where ${ I}_n$ is an $n\times n$ identity matrix.

 \begin{figure} [H]
 \vspace{-2mm}
	\centering
	\includegraphics[width=2in]{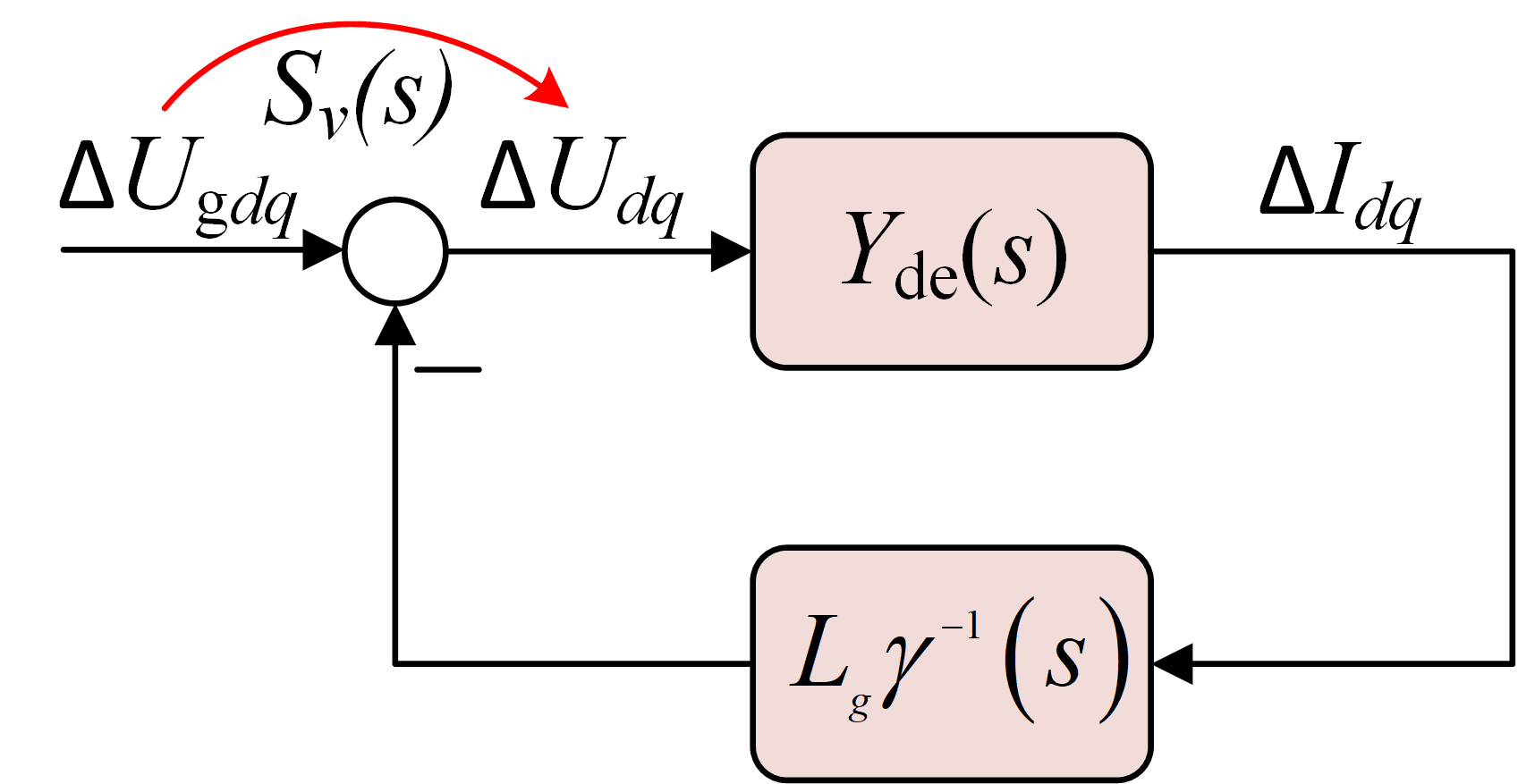}
	\vspace{-3mm}
	\caption{The control diagram of a single converter system.} 
	\vspace{-0.4cm}
	\label{fig3:singleclose}
\end{figure}
\vspace{-4mm}

\subsection{The index for quantifying converter behavior}
As shown in Table~\ref{tab:GFM}, a widely accepted definition of GFM behavior is to maintain a nearly stiff voltage source behavior from sub-transient to transient time frame, which is aligned with roll-off behavior of ``frequency smoothing"~\cite{6:Debryfrequencysmooth,21:grossfrequencysmooth}. {In this paper, this qualitative definition is adopted as the basis for GFM behavior. From the small-signal perspective, converters or control strategies that exhibit such behavior are classified as GFM converters or GFM control, respectively. The ability of a converter to maintain a stiff voltage vector under grid disturbances is used to characterize its voltage source behavior.}

This requires a stiff response in both voltage and frequency dimensions under external grid disturbances, referred to as a {\em ``two-dimensional” voltage source (2D-VS)}  \cite{7:howmany}. In contrast, stiff behavior in only one dimension, either voltage or frequency, is referred to as \textit{``one-dimensional” voltage source (1D-VS)}~\cite{7:howmany}.

Inspired by this qualitative GFM definition, {this section} aims to quantify a converter’s 2D-VS behavior.

{\begin{remark}\label{re1}
Under the small-signal frequency-domain model, frequency and phase are linearly related. At $s=j\omega$, this relationship can be expressed as $\Delta f_u = j\omega \Delta\theta_u/(2\pi)$. Therefore, $\Delta f_u/\Delta f_g = \Delta\theta_u/\Delta\theta_g$, indicating that the stiffness of the angle directly reflects the stiffness of the frequency. Here, $\Delta f_u$, $\Delta f_g$, $\Delta\theta_u$, and $\Delta\theta_g$ denote the converter frequency, grid frequency, converter voltage angle, and grid voltage angle, respectively. Furthermore, under small-signal conditions, the relationship between the dq frame and the polar frame can be expressed as $[\Delta U/U \; ; \; \Delta\theta_u] = U^{-1} e^{J\theta_u}[\Delta U_d \; ; \; \Delta U_q]$, where $\overline{\sigma}(U^{-1}e^{J\theta_u}) \approx 1$ around the nominal operating point. Therefore, the dq-frame sensitivity transfer function $S_v(s)$ captures both voltage magnitude and angle responses. Since the angle response directly reflects frequency dynamics, $S_v(s)$ is adopted to quantify the 2D-VS behavior.

\end{remark}}

As shown in~\eqref{eq3:sensitivity of single VSC}, a “large" ${S}_v(s)$ implies greater voltage deviations under grid disturbances, indicating that the converter operates further from a stiff 2D-VS, {i.e. a weaker ability to maintain a constant voltage vector.} To quantify the magnitude of ${S}_v(s)$, {the \textit{Forming Index} (FI) is introduced. }

\vspace{-2mm}
\begin{definition}[{Forming Index at a given frequency}]\label{def:GFM}
At a frequency $\omega$ of interest, the Forming Index $FI(j\omega)$ is defined as the maximum singular value ($\bar{\sigma}$) of ${S}_v(j\omega)$, i.e., 
\vspace{-1mm}
\begin{equation}\label{eq4:FI}
FI(j\omega)=\bar{\sigma}[{S}_v(j\omega)]\,.
\end{equation}
\end{definition}

$FI(j\omega)$  represents the maximum gain from the grid disturbance voltage vector to the converter voltage vector, i.e.
\vspace{-1mm}
\begin{equation}\label{eq5:meaning of Fi}
\begin{aligned}
FI(j\omega)= \underset{\|\Delta {U}_{{\rm g}dq}(j\omega)\|_2 \neq 0}{{\rm max}} \frac{\|\Delta {U}_{dq}(j\omega)\|_2}{\|\Delta {U}_{{\rm g}dq}(j\omega)\|_2}\,,
\end{aligned}
\end{equation}
where $\|\cdot\|_2$ denotes the $\ell^2$ norm.

As noted in Remark~\ref{re1}, the sensitivity singular values in polar and $dq$ frames differ only by the steady-state constant $U/U_g$. When $U/U_g<1$, the polar-frame formulation can replace the $dq$-frame formulation to avoid misclassifying a simple grounding impedance as GFM. The resulting metric provides an upper bound on~\eqref{eq5:meaning of Fi} and remains compatible with the subsequent analysis when $U/U_g<1$ holds. 

{It should be emphasized that this framework can also be used to quantify the stiffness of the converter output current, which is referred to as current forming behavior~\cite{crossforming}. This can be achieved by evaluating the singular value of the sensitivity function from disturbances to the output current.}
 \begin{figure} [ht]
	\centering
	\includegraphics[width=3.5in]{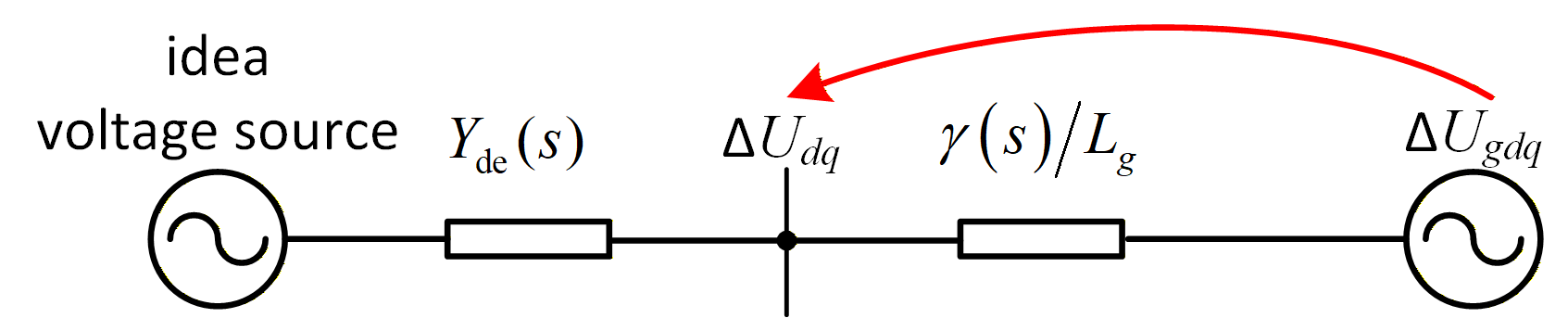}
	\vspace{-3mm}
	\caption{The equivalent circuit of a single converter system} 
	\vspace{-0.2cm}
	\label{fig4:circuit}
\end{figure}

The equivalent circuit of~\eqref{eq3:sensitivity of single VSC} shown in Fig.~\ref{fig4:circuit}, illustrates the division of the voltage vector under grid voltage disturbances, resulting from the interaction between the device and the line admittance. A converter with $FI(j\omega)< 1$ indicates that the converter voltage rejects the grid voltage variations at that frequency. This means the bus voltage behaves more like a stiff 2D-VS. Therefore, $FI(j\omega)< 1$ quantitatively defines GFM capability: a smaller $FI(j\omega)$ corresponds to a stronger GFM capability, with $FI(j\omega)=0$ representing an ideal voltage source. In contrast, $FI(j\omega)\geq1$ indicates that the converter follows or even amplifies the grid voltage variations, which corresponds to GFL behavior.  

As it turns out, the $FI$ also determines the robust stability margin quantifying the distance to instability of the feedback interconnection in Fig.~\ref{fig3:singleclose}.

\begin{lemma}[{$FI$ characterizes the robust stability margin}] \label{robustmargin}
Consider the converter-grid closed loop shown in Fig.~\ref{fig3:singleclose}, the robust stability margin is determined by $\|S_v(s)\|_\infty:=\underset{\forall \omega \in [0,\infty)}{{\rm max}}\bar{\sigma}\left[S_v(j\omega)\right]=\underset{\forall \omega \in [0,\infty)}{{\rm max}}FI(j\omega)$.
\end{lemma}

\begin{proof}
  Consider Fig.~\ref{fig3:singleclose} and let $G(s)=L_{\rm g}\gamma^{-1}(s)$ and $K(s)=Y_{\rm de}(s)$. The stability of a feedback system with the loop transfer function $L(s)=G(s)K(s)$ is determined by the characteristic polynomial ${\rm det}\left[I+L(s)\right]=0$. The system is stable if and only if $L(s)$ satisfies the generalized Nyquist stability criterion. Furthermore, in the open-loop stable case, the robust stability margin is given by the $H_\infty$ norm of the sensitivity function $S(s)=\left[I+L(s)\right]^{-1}$, defined as $\|S(s)\|_\infty:=\underset{\forall \omega \in [0,\infty)}{{\rm max}}\bar{\sigma}\left[S(j\omega)\right]$. A smaller $\|S(s)\|_\infty$ indicates a larger robust stability margin.
\end{proof} 

Based on the qualitative definition of Table~\ref{tab:GFM}, a GFM converter should satisfy the following quantitative conditions: there exist control parameters such that, across a wide range of SCR, $FI<1$ within the frequency band corresponding to the sub-transient to transient time scales.

	\vspace{-3mm}
\subsection{The GFM behavior of different devices}}

The characteristics of converters with different control architectures are analyzed via the proposed index $FI$. The architectures illustrated in Fig. \ref{fig2:control}, include VSG, Droop, VOC~\cite{3:XIUQIANGDVOC}, reactive power control based on PLL (PLL-PQ), PLL-PV, and PLL-VAC \cite{9:pll-gfm}. In addition, the proposed $FI$ is equally applicable to analyzing the 2D-VS behavior of SGs. This can be achieved by modeling the dynamics of a SG as the admittance $Y_{de}(s)$ as defined in~\eqref{eq1:Yvsc}, and then substituting it into~\eqref{eq3:sensitivity of single VSC} and~\eqref{eq4:FI}. In this section, SG cases are also included. {  The SG model~\cite{SG_model} and the base parameters of all devices can be found in Appendix~\ref{sec:model}.}

\begin{figure} 
	\centering
	\includegraphics[width=3.5in]{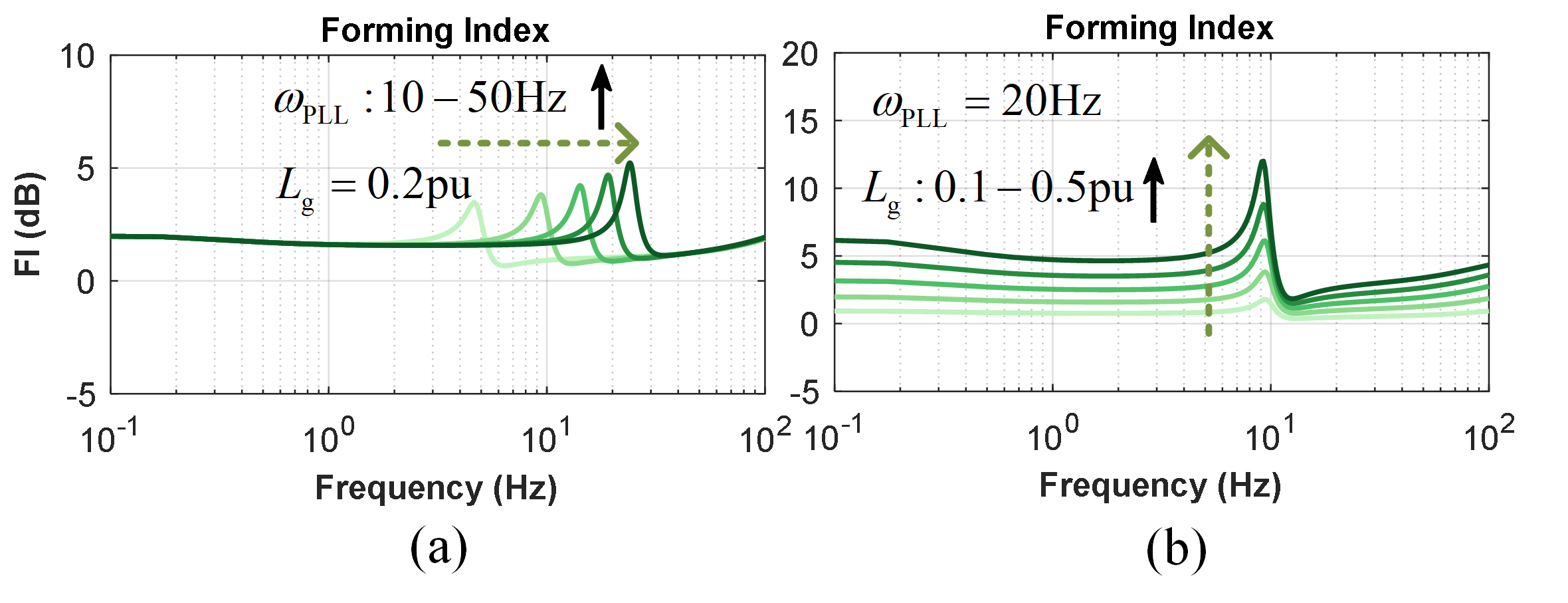}
	\vspace{-7mm}
	\caption{The $FI$s of PLL-PQ. (a) $\omega_{\rm PLL}=10-50$Hz. (b) $L_{\rm g}=0.1-0.5$pu. } 
	\vspace{-0.4cm}
	\label{fig5:PQ-VSC}
\end{figure}

\begin{figure} 
	\centering
	\includegraphics[width=3.5in]{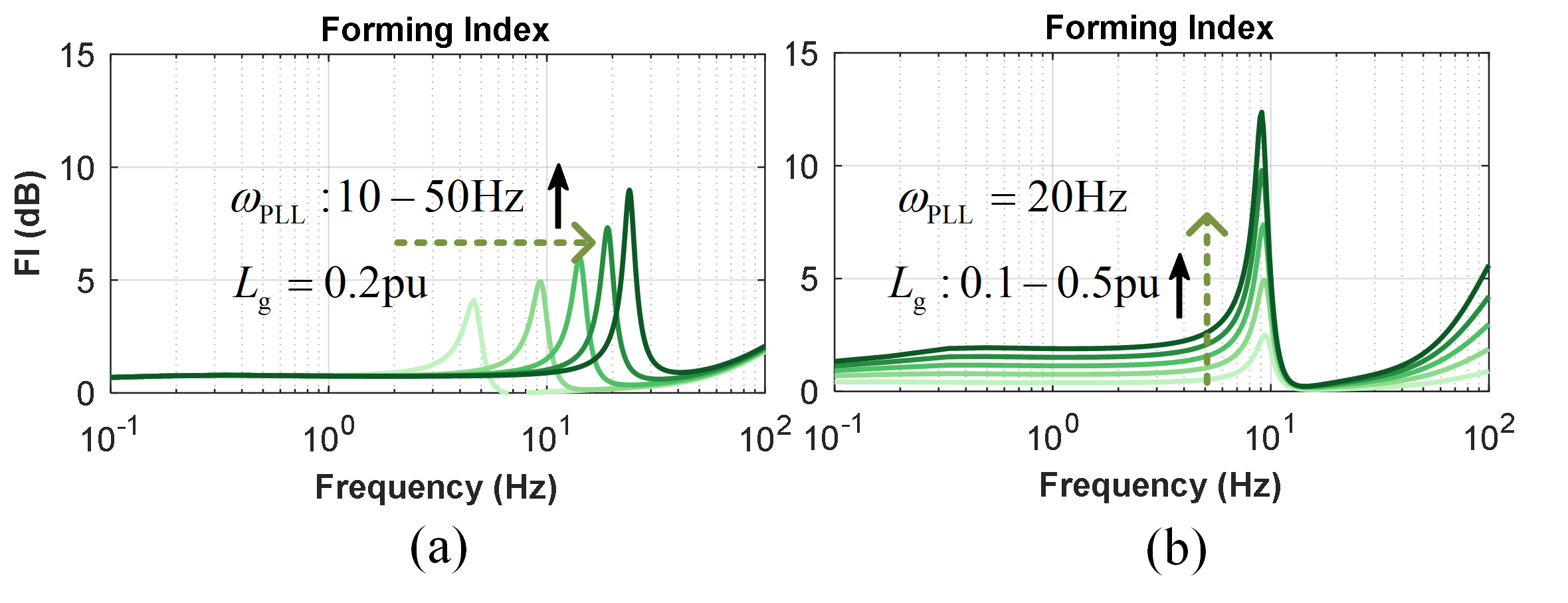}
	\vspace{-7mm}
	\caption{The $FI$s of PLL-PV. (a) $\omega_{\rm PLL}=10-50$Hz. (b) $L_{\rm g}=0.1-0.5$pu. } 
	\vspace{-0.4cm}
	\label{fig6:PLL-PV}
\end{figure}

\begin{example}[{PLL-PQ, and PLL-PV in Fig.~\ref{fig2:control} (a)}] The results of $FI$ for the PLL-PQ and PLL-PV converters with varying PLL bandwidth ($\omega_{\rm PLL}$) and line inductance $L_{\rm g}$ are shown in Figs.~\ref{fig5:PQ-VSC} and \ref{fig6:PLL-PV}, respectively. PLL-PQ and PLL-PV maintain $FI>1$ across the entire frequency range and show a peak at mid-frequencies (approximately aligned with $\omega_{\rm PLL}$), with the peak value increasing as $L_{\rm g}$ increases (i.e. SCR decreases). This indicates that the converter deviates further from a stiff voltage source, with reduced stability margins and an increased risk of resonance, which is consistent with existing research conclusions on the stability of a grid-connected GFL converter~\cite{Linbin:PLL_sy}. 
\end{example}

\begin{example}[VSG, and Droop in Fig.~\ref{fig2:control} (b), {PLL-VAC} in Fig.~\ref{fig2:control} (c), and dVOC in Fig.~\ref{fig2:control} (d), SG]
Corresponding to the sub-transient to transient time scales, a GFM converter {or an SG} is expected to maintain $FI<1$ over the frequency range from a few Hz up to approximately a few hundred Hz, with $FI \approx 0$ desirable in the tens-of-Hz range. Then, the SG and the four control architectures recognized as GFM are examined, with their $FI$ results shown in Figs.~\ref{fig7:VSG and droop} -~\ref{fig11:SG}.

\begin{figure} 
	\centering
	\includegraphics[width=3.5in]{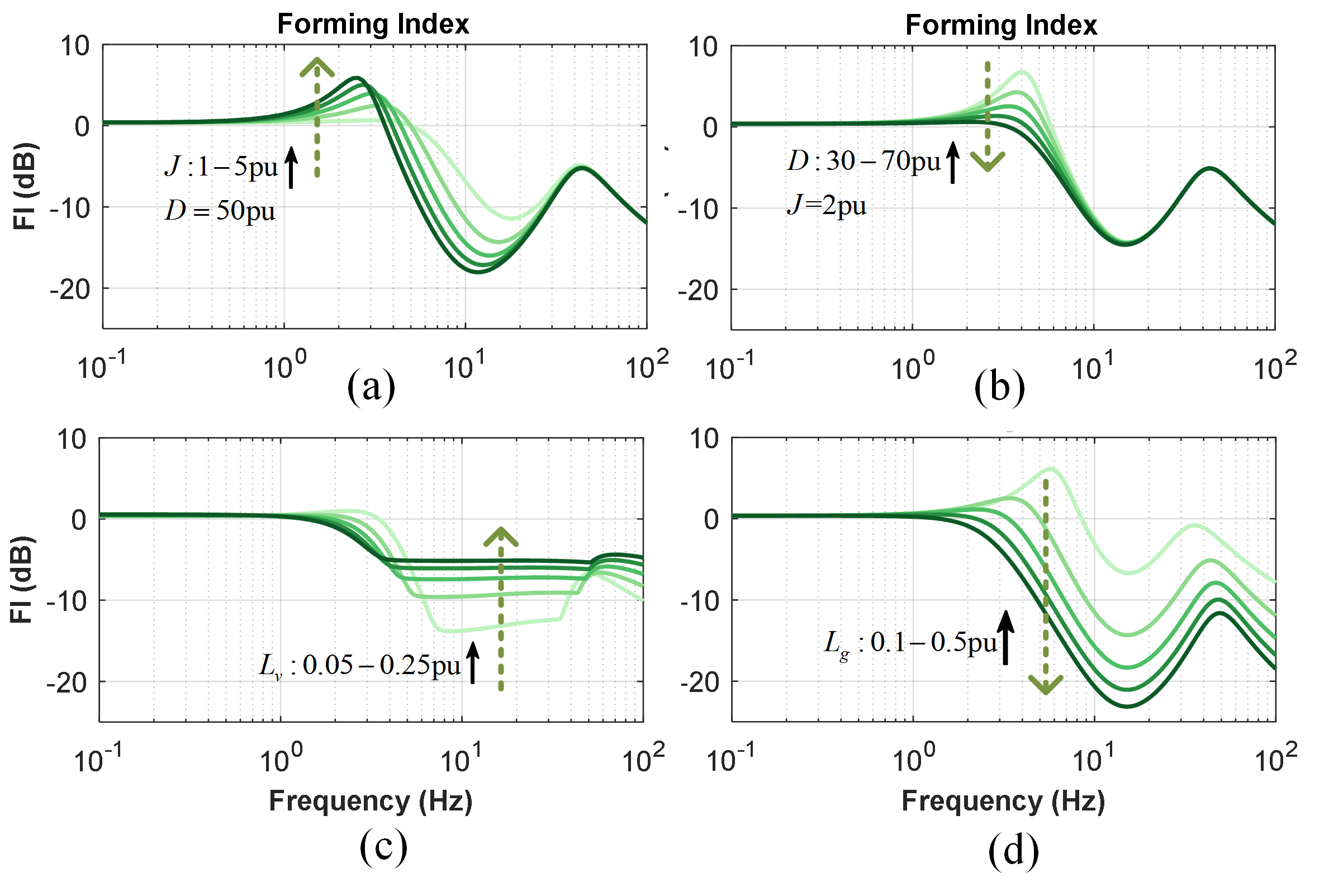}
	\vspace{-7mm}
	\caption{The $FI$s of VSG. (a) $J=1-5$pu, $D=50$pu. (b) $D=30-70$pu. (c) $L_{\rm v}=0.05-0.25$pu. (d) $L_{\rm g}=0.1-0.5$pu.}
	\vspace{-0.4cm}
	\label{fig7:VSG and droop}
\end{figure}
\begin{figure} 
	\centering
	\includegraphics[width=3.5in]{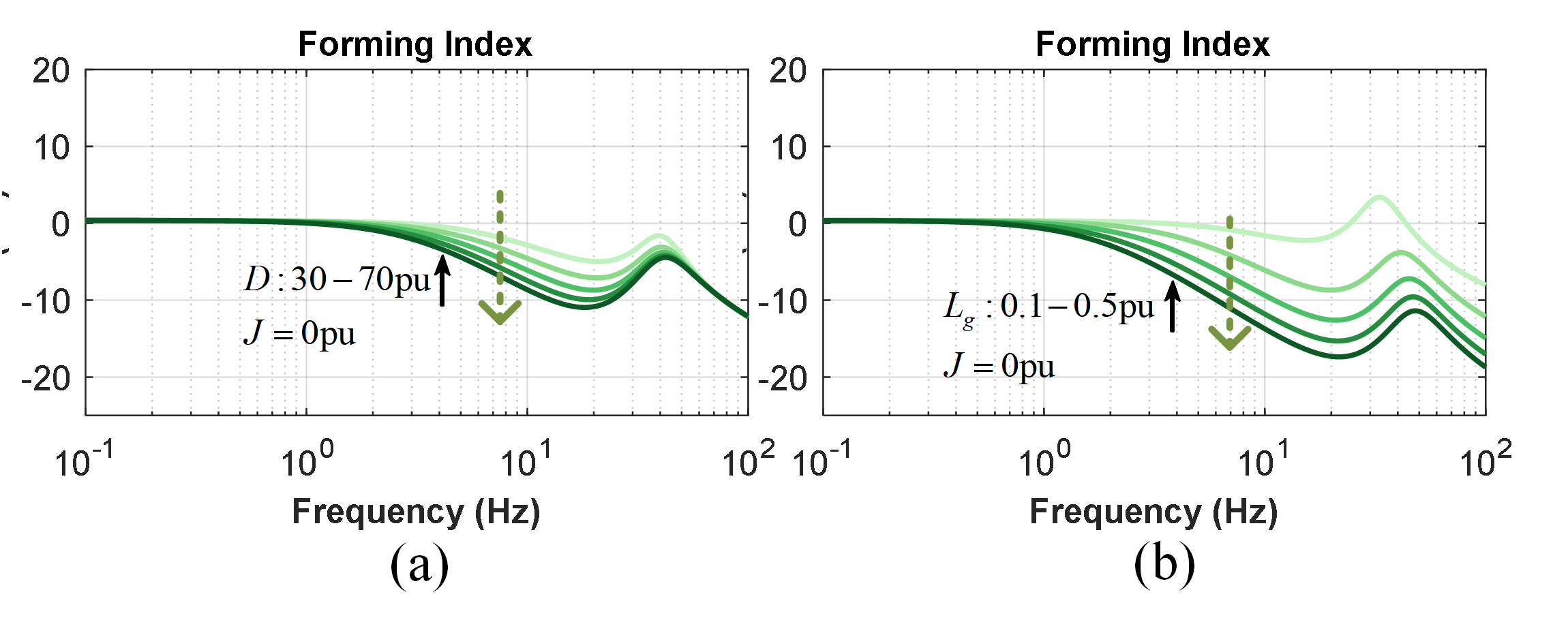}
	\vspace{-7mm}
	\caption{The $FI$s of Droop ($J=0$). (a) $D=1/K_P=30-70$pu. (b) $L_{\rm g}=0.1-0.5$pu.} 
	\vspace{-0.4cm}
	\label{fig8:VSG and droop}
\end{figure}

{\em 1) DC gain and Roll-off behavior.} It is evident that, in Figs.~\ref{fig7:VSG and droop} -~\ref{fig11:SG}, under reasonable parameter settings, the {SG} and four GFM controls have DC gain $FI\approx 1$ ($0$ dB in the plots on a logarithmic scale) at very low frequencies as required for synchronization, and then they exhibit a roll-off behavior with $FI<1$ from a few Hz to a hundred Hz. Nevertheless, careful tuning of control parameters remains essential to ensure that the converter achieves the intended GFM behavior, which is elaborated on below.

{\em 2) Resonance near 50Hz.} A common observation across all four GFM converters {or SG in Figs.~\ref{fig7:VSG and droop} (d) ,~\ref{fig8:VSG and droop} (b),~\ref{fig9:VOC} (a), ~\ref{fig10:PLL-GFM} (b) and~\ref{fig11:SG} }is that, when connected to a high-SCR grid, the $FI$ tends to increase and may exhibit a pronounced peak around 50 Hz~\cite{xiongfei:syn_overview}, necessitating careful tuning of the virtual impedance $Z_v(s)=L_v\gamma^{-1}(s)$ in Fig.~\ref{fig2:control} (b) to suppress this resonance.

{\em 3) Low-frequency peak of VSG, PLL-VAC, {and SG.}} {In Figs.~\ref{fig7:VSG and droop} (a), ~\ref{fig10:PLL-GFM} (b) and~\ref{fig11:SG}, a GFM converter or an SG with inertia $J$} may exhibit peaks at low frequencies (around 5Hz aligned with the bandwidth of synchronization control). For GFM converters, although increasing the inertia can accelerate the decline of $FI$ in the high frequency range, it may also worsen the low frequency GFM behavior unless it is carefully paired with a virtual impedance (Fig.~\ref{fig7:VSG and droop} (d)) or a large damping coefficient (Fig.~\ref{fig7:VSG and droop} (c)). 
{ For an SG, careful tuning of Power System Stabilizer (PSS) parameters is typically required to suppress low-frequency peaks. The phenomena 2) and 3) align with two common types of oscillations \cite{xiongfei:syn_overview}: low-frequency and synchronization oscillations.}

{\em 4) Low-frequency peak of dVOC.} As shown in Fig. \ref{fig9:VOC}, due to the inherent voltage oscillatory behavior in dVOC, increasing the reactive power droop coefficient $K_Q$ results in a higher $FI$ around $1$Hz. Within the reactive power control bandwidth, the voltage magnitude changes in response to reactive power variations. Nevertheless, the bandwidth is typically set outside the sub-transient to transient time scales, and therefore does not affect the GFM behavior that is truly expected.

{\em 5) { GFM behavior of PLL-VAC.}} As shown in Fig.\ref{fig10:PLL-GFM}, with properly tuned parameters, {PLL-VAC} can also achieve GFM behavior, thus proving that relying solely on synchronization control to distinguish between GFM and GFL behaviors is insufficient at least from the small-signal perspective. The virtual admittance is crucial for enabling GFM behavior in {PLL-VAC}. When connected to a high SCR grid, an insufficient virtual admittance can cause $FI$ to increase monotonically above $1$ in the high-frequency range. 

 \begin{figure} 
	\centering
	\includegraphics[width=3.5in]{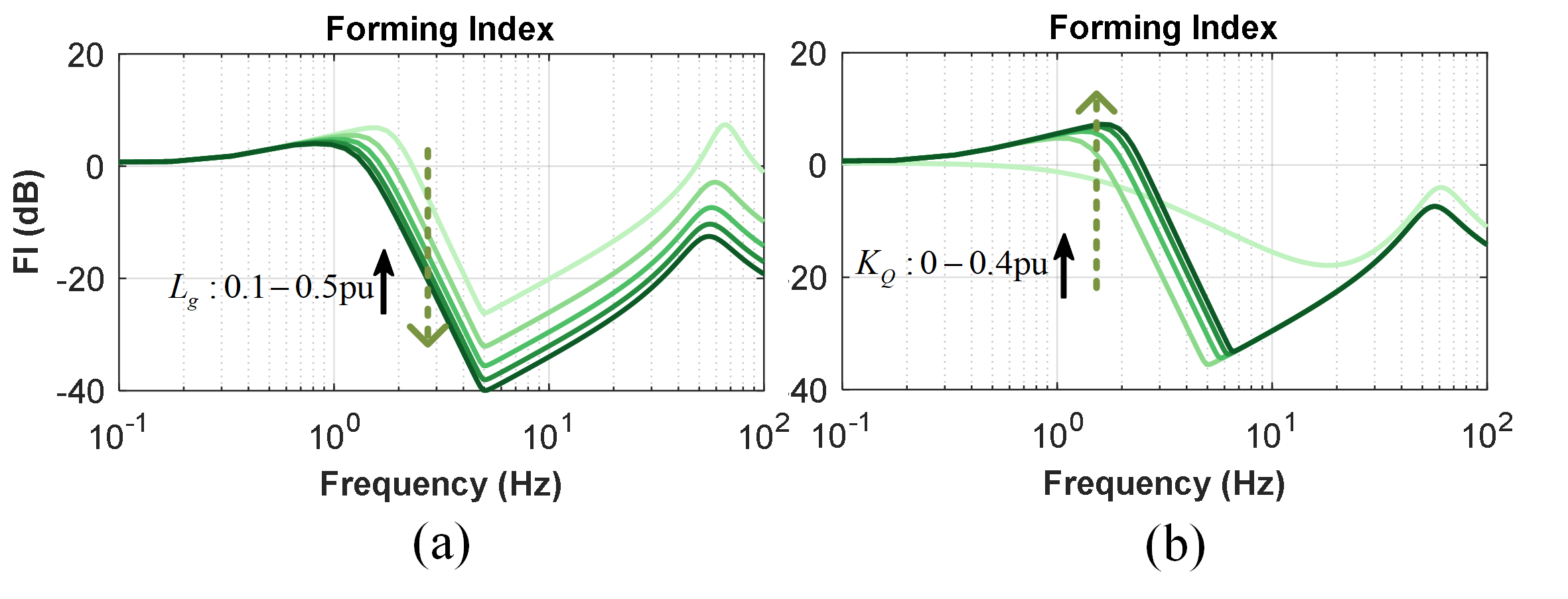}
	\vspace{-7mm}
	\caption{The $FI$s of dVOC. (a) $L_{\rm g}=0.1-0.5$pu. (b) $K_{Q}=0-0.4$pu. } 
	\vspace{-0.4cm}
	\label{fig9:VOC}
\end{figure}
  \begin{figure} 
	\centering
	\includegraphics[width=3.5in]{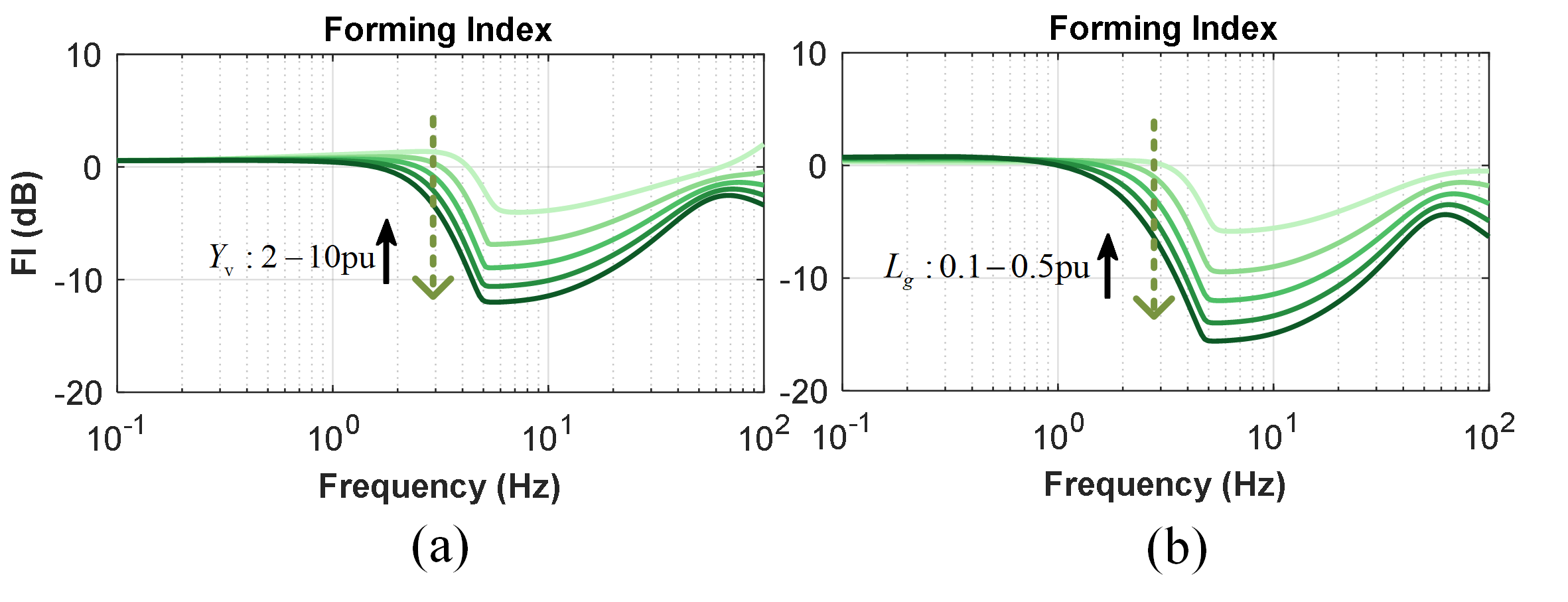}
	\vspace{-7mm}
	\caption{The $FI$s of {PLL-VAC}. (a) $Y_{\rm v}=2-10$pu. (b) $L_{\rm g}=0.1-0.5$pu.}
	\vspace{-0cm}
	\label{fig10:PLL-GFM}
\end{figure}
  \begin{figure} 
	\centering
	\includegraphics[width=3.5in]{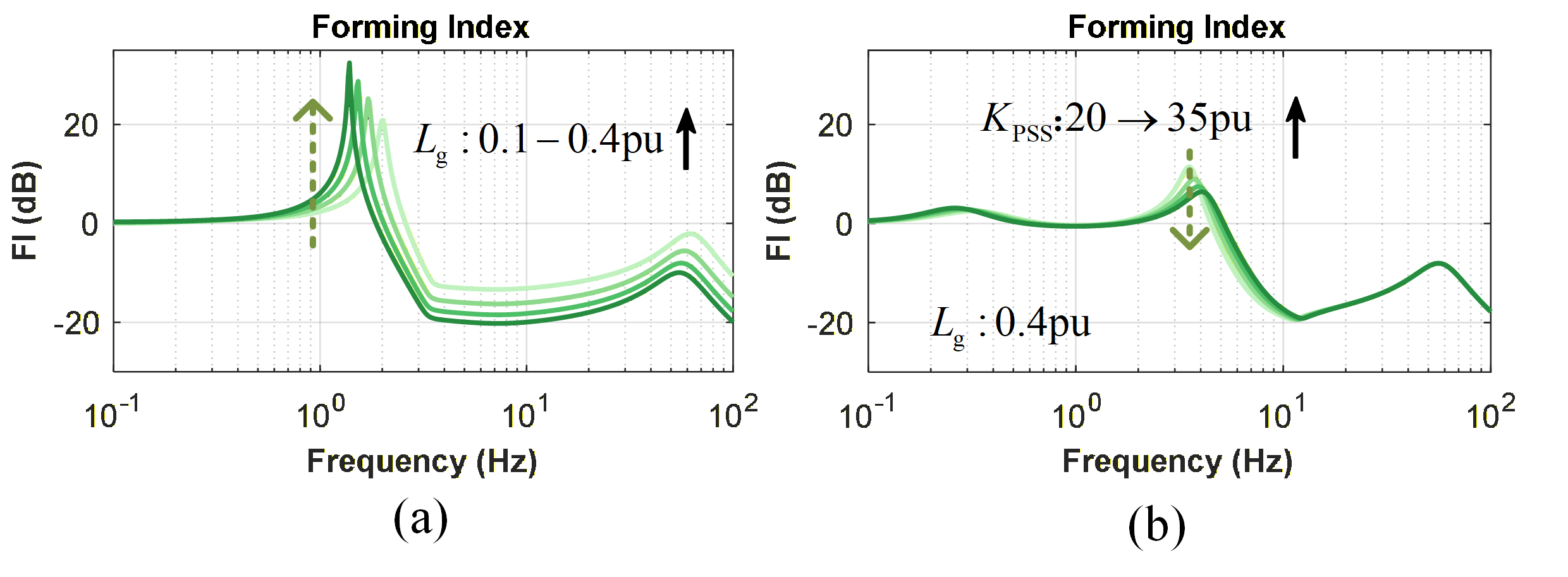}
	\vspace{-7mm}
	\caption{{The $FI$s of SG. (a) $L_{\rm g}=0.1-0.4$pu, without PSS control. (b) $K_{\rm PSS}=20-35$pu (PSS gain parameter), with PSS control.}}
	\vspace{-0cm}
	\label{fig11:SG}
\end{figure}

\end{example}

In summary, whether a PLL is employed cannot be the sole criterion for GFM behavior. A key characteristic of GFM is that, through appropriate parameter tuning, it can maintain $FI<1$ from a few Hz to around a few hundred Hz and exhibit a roll-off trend under any grid condition. 

\subsection{Comparison and discussion   }
This subsection compares $FI$ with existing indicators, such as the converter impedance norm (IN)~\cite{7:howmany}, the frequency smoothing (FS) metrics~\cite{21:grossfrequencysmooth} and the passivity of converter admittance, which are shown in Table~\ref{T2:GFM_metrics}. In the definition, $Z_{\rm de}(s):=Y^{-1}_{\rm de}(s)$ of~\eqref{eq1:Yvsc}, $G_{f_{\rm g},f_u}(s)$ is the transfer function from grid frequency $f_g$ to converter frequency $f_u$.

{\em 1) Converter impedance norm.} A converter can be considered truly only stiff when $\overline{\sigma}\left[{Z}_{\rm de}(j\omega)\right]\approx0$ , and no rigorous criteria exist beyond this. Hence, the impedance norm alone lacks a clear threshold for distinguishing GFM behavior and is therefore an insufficient metric. 

{\em 2) Frequency smoothing.} Because ${\Delta f_u}/{\Delta f_g}={\Delta \theta_u}/{\Delta \theta_g}$, $G_{f_{\rm g},f_u}(s)$ is the bottom-right element of $U^{-1}e^{J(\theta_u-\theta_g)}{S}_{v}(s)$ (transformation from the $dq$ frame to the polar coordinate frame), and it is thus naturally bounded by our index, such that $\left|G_{f_{\rm g},f_u}(j\omega)\right|\leq U^{-1}FI(j\omega)$. This metric only describes the frequency stiffness under the grid frequency disturbance, but it does not fully capture the 2D-VS behavior.

{{\em 3) Passivity.} Under small-signal conditions, if $Y_{\rm de}(s)$ is stable, the condition ${\rm Re}(Y_{\rm de}(j\omega))\geq0$ is typically referred to as passivity or positive realness~\cite{passiveofgfm}. At the device level, the relationship between $FI$ and passivity is as follows.

\begin{lemma}[Relationship between $FI$ and device passivity] \label{FIandpassivity}
Consider the single converter grid connected system in Fig.~3 and define
$H(s):=\gamma^{-1}(s)Y_{\rm de}(s)$. At a given frequency $\omega$, if $H(j\omega)$ is strictly positive real (satisfies passivity at this frequency), i.e.,
\[
H(j\omega)+H^H(j\omega)\succ 0 \,,
\]
then $FI(j\omega)<1$. Further, passivity at $\omega$ is also necessary for $FI(j\omega)<1$ in the limit as $L_g \to 0$.
\end{lemma}
\begin{proof}
    See Appendix~\ref{sec:appendixA}.
\end{proof}

In fact, $\gamma^{-1}(s)$ can be interpreted as an inductive multiplier that maps the direction associated with the inductive network onto the positive-real axis, while characterizing the similarity between the device and network admittances. The resulting property is termed inductive-multiplier passivity, a special case of multiplier passivity.

As noted in~\cite{passiveofgfm}, the condition $\mathrm{Re}(Y_{\rm de}(j\omega))>0$ is difficult to satisfy in the low-frequency range and is thus usually imposed only above 3 Hz. In comparison, condition $\mathrm{Re}(H(j\omega))>0$ can be achieved at low frequencies, thereby covering a wider frequency band.

}

Compared to the aforementioned metrics, the proposed index incorporates both voltage and frequency behavior and also provides a clear boundary ($FI<1$) for GFM behavior. 

Last, considering $FI$ and power-frequency droop: $FI \approx 0$ signifies that the converter frequency fully resists grid frequency variations. A stiff voltage source with a constant frequency is thus perfectly GFM behavior, but it provides no power–frequency droop. In addition, the $FI$ makes no statement whether droop control is provided at low frequencies ($s\approx0$). This must be assessed through the transfer function from grid frequency $f_g$ to converter active power $P$, as discussed in Ref.~\cite{21:grossfrequencysmooth}.

\begin{table}
	\vspace{0mm}
    \centering
        \caption{Different metrics for qualifying voltage source behavior}
        \label{T2:GFM_metrics}
        \renewcommand{\arraystretch}{1.8} 
    \begin{tabular}{|c|c|c|c|}\hline
         FI&  IN~\cite{7:howmany}& FS~\cite{21:grossfrequencysmooth} &{ Passivity}\\\hline
          $\overline{\sigma}\left[S_v(j\omega)\right]$&  $\overline{\sigma}\left[{Z}_{\rm de}(j\omega)\right]$& $\left|G_{f_{\rm g},f_u}(j\omega)\right|$ & {${\rm Re}(Y_{\rm de}(j\omega))$}\\\hline
    \end{tabular}

    \label{tab:placeholder}
\end{table}

\section{System Level: System Strength} \label{sec:SS}
\subsection{Small signal power system modeling}
Consider a power system consisting of $n$ devices (SG, converter, load, etc.) connected at buses $ \left\{1,\dots,n \right\} $ , as well as $m$ interior buses $ \left\{n+1,\dots,n+m \right\} $ and a common grounded bus $\left\{n+m+1\right\}$. Each device can be modeled as an admittance transfer function matrix $Y_{\rm de}(s)\in  \mathcal{RH}^{2 \times 2}_\infty$, similar to~\eqref{eq1:Yvsc}. The dynamics of the $n$ devices are,
\vspace{-1mm}
\begin{equation}\label{eq6:ndevice}
      \Delta {\bm I}_N = {\bm Y}^N_{\rm de}(s)
        \Delta {\bm U}_N\,,
\end{equation}
where ${\bm Y}^N_{\rm de}(s):={\rm diag}\left\{Y_{\rm de,1}(s),\cdots, Y_{\rm de,n}(s) \right\}$ is the block-diagonal matrix of admittances normalized by the converters’ rated capacities, $ \Delta\bm{U}_N={\bm S}^{1/2}_{\rm B,N}\begin{bmatrix} \Delta U^{\top}_{dq,1}  & \dots   & \Delta U^{\top}_{dq,n} \end{bmatrix}^{\top }$ and $ \Delta\bm{I}_N={\bm S}^{1/2}_{\rm B,N}\begin{bmatrix} \Delta I^{\top}_{dq,1}  & \dots   & \Delta I^{\top}_{dq,n} \end{bmatrix}^{\top }$ are the voltage and current vectors at buses $ \left\{1,\dots,n \right\} $  respectively, both scaled by the same capacity matrix ${\bm S}^{1/2}_{\rm B,N}$ to transfer capacity differences to the network side. $ {\bm S}_{{\rm B},N}:={\rm diag}\left\{S_{{\rm B},1},\cdots, S_{{\rm B},n} \right\}\otimes I_2$ is the block-diagonal matrix of converter ratings.

Next, the transfer function matrix of a power network is derived. Consider a general setting where the dynamics between bus $i$ and bus $j$ are given by
\vspace{-1mm}
\begin{equation}\label{eq7:ijline}
    \begin{bmatrix}
    \Delta I_{d,ij}\\
    \Delta I_{q,ij}
    \end{bmatrix}= {Y}_{ij}(s) \left(\begin{bmatrix}
    \Delta U_{d,i}\\
    \Delta U_{q,i}
    \end{bmatrix}- \begin{bmatrix}
    \Delta U_{d,j}\\
    \Delta U_{q,j}
    \end{bmatrix}\right)\,,
\end{equation}
where,  $\begin{bmatrix}
    \Delta I_{d,ij}&
    \Delta I_{q,ij}
    \end{bmatrix}^\top$ is the current vector from bus $i$ to bus $j$ in the global dq frame, and ${Y}_{ij}(s)$ is the $2\times2$ admittance transfer function matrix between bus $i$ and bus $j$. For example, a line between bus $i$ and bus $j$ composed of a resistor and an inductor can be written as
 \vspace{-1mm}
\begin{equation}\label{eq8:Yline}
{Y}_{ij}(s)=B_{ij}{ \gamma}_{ij}(s)=B_{ij}{\begin{bmatrix}
  \frac{s}{\omega_0}+\tau_{ij}  & -1 \\ 1 & \frac{s}{\omega_0}+\tau_{ij}
\end{bmatrix}^{-1}}\,,
\end{equation}
{and the shunt capacitor at bus $i$ (with $j=n+m+1$) can be expressed as
\begin{equation}\label{eq8:YC}
{Y}_{ij}(s)=C_{i}\begin{bmatrix}
  \frac{s}{\omega_0}  & -1 \\ 1 & \frac{s}{\omega_0}
\end{bmatrix}\,,
\end{equation}
where $B_{ij}$ is the line susceptance between bus $i$ and bus $j$, $\tau_{ij}$ is the resistance to inductance ratio, and $C_i$ is the capacitance at bus $i$. }Then the $2(n+m)\times2(n+m)$ admittance matrix of the network ${\bm Y}(s)$ is
 \vspace{-1mm}
\begin{equation}\label{eq9:Ylin}
{\bm Y}(s):=\begin{bmatrix} {\bm Y}_1 (s) \in  \mathcal{RH}^{2n \times 2n}_\infty& {\bm Y}_2 (s)\in  \mathcal{RH}^{2n \times 2m}_\infty\\{\bm Y}_3 (s)\in  \mathcal{RH}^{2m \times 2n}_\infty & {\bm Y}_4 (s)\in  \mathcal{RH}^{2m \times 2m}_\infty\end{bmatrix}\,,
\end{equation}
where the blocks of ${\bm Y}(s)$ are
 \vspace{-1mm}
\begin{equation}\label{eq10:blockofYnet}
\begin{aligned}
   {\bm Y}_{ij}(s)=-Y_{ij}(s),i\ne j\,,\\ 
   {\bm Y}_{ii}(s)= {\textstyle \sum_{j=1, j\neq i}^{n+m+1}}  Y_{ij}(s)\,.
\end{aligned}
\end{equation}

{ It is worth noting that SGs can also be treated as a device in the framework of~\eqref{eq6:ndevice}. The case study involving SGs is presented in Section V-B. If the SG swing dynamics and internal electromotive-force dynamics are neglected, the SG can be approximated by its sub-transient inductance, represented in the same form as~\eqref{eq8:Yline}, and absorbed into the network model.}

\begin{figure} 
	\centering
	\includegraphics[width=3.1 in]{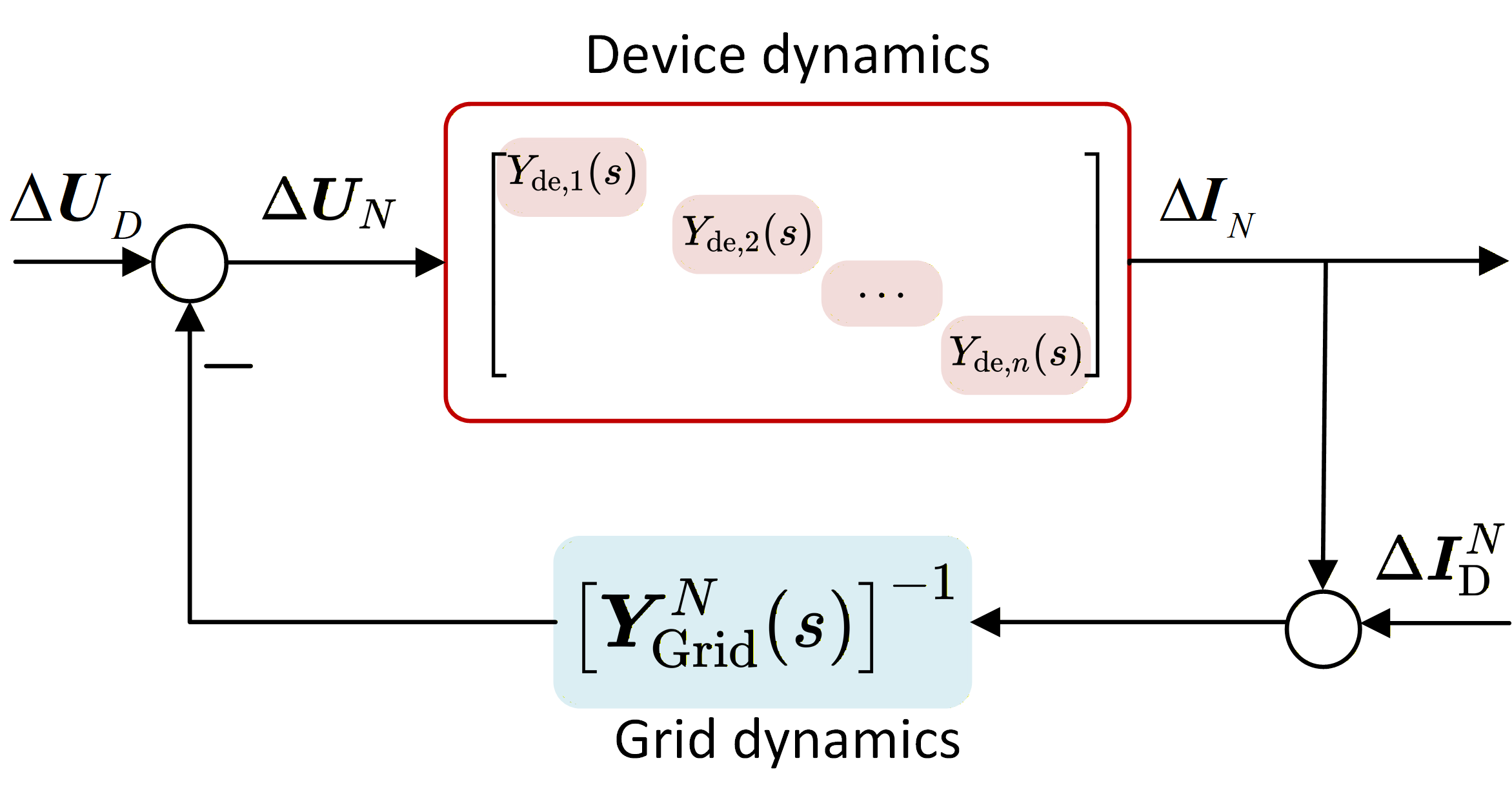}
	\vspace{-3mm}
	\caption{The control diagram of a power system.} 
	\vspace{-0.4cm}
	\label{fig11:muilti-diagram}
\end{figure}

By eliminating the $m$ interior buses by Kron reduction, {a reduced $2n\times 2n$ network matrix $    \Delta {\bm I}_N+\Delta {\bm I}^N_{\rm D} ={\bm Y}^N_{\rm Grid}(s)\Delta {\bm U}_N $ is obtained based capacity normalization,}
\begin{equation}\label{eq11:Ygrid}
    {\bm Y}^N_{\rm Grid}(s):={\bm S}^{-1/2}_{\rm B,N}\left({\bm Y}_1 (s)-{\bm Y}_2(s){{\bm Y}_4} ^{-1}(s){\bm Y}_3(s)\right){\bm S}^{-1/2}_{\rm B,N}\,.
\end{equation}
As shown in Fig.~\ref{fig11:muilti-diagram}, Eqs.~\eqref{eq6:ndevice} and~\eqref{eq11:Ygrid} constitute the closed-loop dynamics of the power system. Consider $\Delta {\bm I}^N_{\rm D}={\bm S}^{1/2}_{\rm B,N}\begin{bmatrix} \Delta I_{{\rm D}dq,1} &\cdots &\Delta I_{{\rm D}dq,n} \end{bmatrix}^{\top}$ as a disturbance current vector injected into buses $\left\{1,\cdots,n\right\}$, and after eliminating $\Delta {\bm I}_N$, the dynamics of the buses can be expressed as
\vspace{-1mm}
\begin{equation}\label{eq12:additional}
\begin{aligned}
         \Delta {\bm U}_{N}&=\left[ {\bm Y}^N_{\rm Grid}(s)+{\bm Y}^N_{\rm de}(s)\right]^{-1}\Delta {\bm I}^N_{\rm D} \\
        &=:{ {\bm Y}^{-1}_{\rm Cl}}(s)\Delta {\bm I}^N_{\rm D}\,,
\end{aligned}
\end{equation}
 where  ${ {\bm Y}^{-1}_{\rm Cl}}(s)=:{\bm Z}_{\rm Cl}(s)$ is the {\em sensitivity transfer function} from current disturbance to voltage vector, which is equivalent to the system's closed-loop impedance matrix.

\vspace{-3mm}

{\subsection{The concept of system strength}}
A necessary requirement for a stable power system is to ensure that the multi-bus voltage vectors remain within safe limits under a disturbance, i.e., the system has stiff voltages at all buses. To formalize these requirements, CIGRE \cite{10:cigre} and AEMO \cite{strength:aemo} have introduced the concept of {\em system strength}. It is defined as “the ability of the power system to maintain and control the voltage waveform at given any location, both during steady-state operation and following a disturbance". {Steady state reflects static strength, while under disturbances reflects dynamic strength~\cite{strength_magazine}. Moreover, depending on the magnitude of the disturbance, strength should be studied separately from the perspectives of small-signal and large-signal~\cite{strength_magazine}. This paper focuses on the small-disturbance perspective.}

As derived in~\eqref{eq12:additional}, the voltage waveform response is determined by the closed-loop dynamics formed by the power network and the devices. Although existing studies have proposed various SCR–based indices to quantify system strength, they typically consider only the power network, which fails to accurately capture the closed-loop voltage response. 

Inspired by the above qualitative definition of system strength, this section proposes a quantitative metric evaluating the voltage offset under current (and power) disturbances.

\begin{definition}[{System Strength}]\label{def:SS}
Consider the interconnected device and grid system in~\eqref{eq12:additional}, at a given frequency $\omega$, the system strength $\kappa(j\omega)$ is mathematically defined as the worst sensitivity
\vspace{-1mm}
\begin{equation}\label{eq13:SS}
\begin{aligned}
  \kappa(j\omega):&=\overline{\sigma}^{-1}[{{\bm Z}}_{\rm Cl}(j\omega)]=\underline{\sigma}[{{\bm Y}}_{\rm Cl}(j\omega)]\\
&=\underline{\sigma}[{{\bm Y}}^N_{\rm Grid}(j\omega)+{\bm Y}^N_{\rm de}(j\omega)]\\
& = \left({  \underset{\|\Delta {\bm I}^N_{{\rm D}}(j\omega)\|_2 \neq 0}{{\rm max}}\frac{\|\Delta {\bm U}_N(j\omega)\|_2} {\|\Delta {\bm I}^N_{{\rm D}}(j\omega)\|_2}}\right)^{-1}\,.  
\end{aligned}
\end{equation}
\end{definition}

    According to~\eqref{eq13:SS}, when a sinusoidal current or power disturbance with frequency $\omega$ is injected, $1/\kappa(j\omega)$ quantifies the gain from the current or power disturbance to the voltage. Thus, a larger $\kappa$ corresponds to smaller voltage variations under disturbances, indicating stronger disturbance rejection.

Similar to $FI$, the system strength defined via singular value is also compatible with the $\mathcal{H}_\infty$ norm of the sensitivity function, i.e., a higher system strength thus corresponds to a larger robust stability margin. 

According to empirical insights from robust stability margins~\cite{skogestad2005multivariable}, the system is considered {\em very weak} when $\kappa<0.5$, which poses a higher risk of instability or voltage limit violations. A power system is regarded as {\em weak} when $0.5<\kappa<1$, and as {\em strong} when $\kappa>1$.

  The derivations from~\eqref{eq6:ndevice} to~\eqref{eq12:additional} can be equivalently reformulated in the power coordinate by replacing current with power, yielding a power to voltage sensitivity transfer function ${\bm U}^N_0{ {\bm Y}^{-1}_{\rm Cl}}(s)$ analogous to~\eqref{eq12:additional}, where ${\bm U}^N_0={\rm diag}\left\{ \left[\begin{smallmatrix}
    U_{d,i} & U_{q,i} \\ U_{q,i} & -U_{d,i}
\end{smallmatrix}\right]\right\} ,i=1,\cdots,n$. At the equilibrium point, every diagonal block has $\sigma({\bm U}^N_{0,i})\approx 1$. Therefore, the same system strength definition $\kappa$ applies under both power disturbances and current disturbances {under the small signal condition.}

\begin{remark}
   \label{remark:single and multi}
    Applying the same capacity scaling ${\bm S}^{1/2}_{{\rm B},N}$ to both the current input and voltage output renders homogeneous multi-device systems with uniform $\tau$ equivalent to a single-device system without affecting the quantification of the voltage response. {Specifically, for ${Y}_{{\rm de},1}(s)=\cdots={Y}_{{\rm de},n}(s)$, ignoring the capacitor effect, since ${\bm B}^N_{\rm Grid}={\bm S}^{-1/2}_{{\rm B},N}{\bm B}_{N}{\bm S}^{-1/2}_{{\rm B},N}$ is Hermitian, which is the static network admittance matrix based capacity normalization, and ${{\bm Y}}^N_{\rm Grid}(s)={{\bm B}}^N_{\rm Grid}\otimes \gamma(s)$, then $\underline{\sigma}[{{\bm Y}}_{\rm Cl}(j\omega)]=\underset{\forall i}{{\rm min}}\,\sigma\left[{Y}_{{\rm de},i}(j\omega)+\sigma_i\gamma(j\omega)\right]$ is obtained. Here, $\sigma_i$ is the $i$-th singular value of ${{\bm B}}^N_{\rm Grid}$ and is equivalent to the SCR of a single device system.}
\end{remark}

System strength is jointly determined by the device ${\bm Y}^N_{\rm de}(s)$ and the power grid ${\bm Y}^N_{\rm Grid}(s)$. Based on this, grid strength is distinguished here: after ignoring the dynamics of the devices ${\bm Y}^N_{\rm de}(s)$, the strength of the grid is referred to as grid strength.

\begin{definition}[{Grid strength}]
\label{def:GS}
    At a given frequency $\omega$, the grid strength $\alpha(j\omega)$ is defined as 
   \begin{equation}\label{eq15:grid strength}
\begin{aligned}
  \alpha(j\omega):=\underline{\sigma}\left[ {{\bm Y}}^N _{\rm Grid}(j\omega)\left({I}_n \otimes {\gamma^{-1}_0}(j\omega)\right)\right]\,.
\end{aligned}
\end{equation}
where $\gamma_0(s)$ is as in~\eqref{eq2:Yline} with a constant $\tau$.
\end{definition}

By extracting $\gamma^{-1}_0(s)$, the resulting grid strength metric can be made compatible with the SCR derived from a static grid network matrix, which is discussed in the following section.

In this concept, SCR metrics are used to evaluate grid strength rather than system strength. This distinction enables a separate assessment of grid-side influences and underpins grid code requirements for device interconnection. For example, devices are typically required to ensure stable operation in grids with SCR larger than 1.5.

\begin{proposition}
\label{grid strength}
  The relationship between system strength and grid strength is given by:  $\kappa(j\omega)\geq\underline{\sigma}[{\gamma}(j\omega)]\alpha(j\omega)-\bar{\sigma}[{\bm Y}^N_{\rm de}(j\omega)]$, that is, enhancing grid strength can improve a lower bound for system strength. 
\end{proposition}
\begin{proof}
     This conclusion follows directly from $\underline{\sigma}(A+B)\geq\underline{\sigma}(A)-\overline{\sigma}(B)$, as outlined in the Introduction.
          \vspace{-2mm}
\end{proof}

In addition, this subsection also assesses the strength at each individual bus to identify the weak points for precise enhancement. 

\begin{definition}[{Bus strength}]
\label{def:BS}
    At a given frequency $\omega$, the bus strength of bus $i$ is defined as
    
   \begin{equation}\label{eq16:bus strength}
\begin{aligned}
  \kappa^{-1}_i(j\omega)= {\rm max}\left({\textstyle \sum_{j=1}^{n}}\overline{\sigma}({{\bm Z}}_{{\rm Cl},ij}(j\omega)),\,{\textstyle \sum_{j=1}^{n}}\overline{\sigma}({{\bm Z}}_{{\rm Cl},ji}(j\omega))\right),
\end{aligned}
\end{equation}
where ${{\bm Z}}_{{\rm Cl},ij}(s)$ is the $2\times2$ block matrix in the $i$-th row and $j$-th column of ${{\bm Z}}_{\rm Cl}(s)$.
\end{definition}

Bus strength lower-bounds system strength, i.e., if all buses meet the threshold, then $\kappa(j\omega)\geq {\rm min}(\kappa_i(j\omega))$ guarantees the system strength,  which can be proved by applying the singular value inequalities in the Introduction. 

{ The proposed index is general and can be used to assess strength for networks that include shunt capacitors and lines with different resistance-to-inductance ratios. In the following analysis, to obtain analytical insights, the following assumptions are adopted: a uniform resistance-to-inductance ratio and neglecting the effect of shunt capacitors.}

\vspace{-2mm}
{\subsection{Discussion of different system strength metrics}}
This subsection will review, discuss and compare three strength metrics to equivalent SCR (ESCR) proposed by CIGRE \cite{10:cigre},  generalized SCR (gSCR) \cite{Huanhai:gSCR} and {\em system passivity}~\cite{xiongfei:passive}, which are detailed in Table~\ref{T3:strength_metrics}. Here, ${S}_{{\rm B},i}$ is the capacity of the $i$-th device,  $Z_{ij}$ is the element of ${\bm B}^{-1}_N$.
\begin{table}
    \centering
        \caption{Different metrics for qualifying strength}
        \label{T3:strength_metrics}
        \renewcommand{\arraystretch}{1.8} 
    \begin{tabular}{|c|c|}\hline
         Metrics &  Definition\\\hline
          \makecell{${\rm ESCR}_i$~\cite{10:cigre}}&   \makecell{$\frac{1}{{\textstyle \sum_{j=1}^{n}}S_{{\rm B},j}|Z_{ij}|}$}\\\hline
 gSCR~\cite{Huanhai:gSCR}&$\underline{\lambda}\left({\bm S}^{-1}_{{\rm B},N}{\bm B}_{N}\right)$\\\hline
 Passivity~\cite{xiongfei:passive}&$p(j\omega):={\rm Re}({\bm Y}_{\rm cl}(j\omega))$\\ \hline\end{tabular}
\vspace{-5mm}
\end{table}

{\em 1) gSCR}. Considering grid line dynamics in~\eqref{eq8:Yline} with identical $\tau$, then obtain $\alpha=\underline{\sigma}({{\bm B}}^N_{\rm Grid})={\rm gSCR}$, which is a special case of the proposed grid strength in~\eqref{eq15:grid strength}. { Therefore, it is more reasonable to extend the definition of gSCR to~\eqref{eq15:grid strength}. }

\begin{remark}
For a single GFL converter, a lower SCR indicates reduced stability, and the SCR at the stability boundary is termed the critical SCR (CSCR)~\cite{Huanhai:gSCR}. As noted in Remark~\ref{remark:single and multi}, when converters are homogeneous, the gSCR of a multi-converter power system equals the SCR of a single-converter system. Thus, the system stability margin can be quantified as  $\frac{{\rm gSCR}-{\rm CSCR}}{\rm CSCR}$, effectively decoupling device and grid impacts. 
\end{remark}

{\em 2) ESCR (also known as MSCR)}. The physical meaning of ${\rm ESCR}_i$ is the short-circuit current at bus $i$, which reflects the steady-state voltage deviation of bus $i$ under current disturbances. In addition, $1/{\rm ESCR}_i$ is the sum of the $i$-th row of matrix $\left({{\bm S}^{-1}_{{\rm B},N}{\bm B}_N}\right)^{-1}$. Accordingly, $1/{\rm gSCR}\leq {\rm max} (1/{\rm ESCR}_i)$, i.e. ${\rm gSCR}\geq {\rm min}\,{\rm ESCR}_i$, which indicates that ESCR is more conservative.

 {\em 3) Passivity}. { Passivity serves as a tool not only for evaluating device behavior but also for assessing the closed-loop stability of the system.} When ${{\bm Y}}_{\rm cl}(s)$ contains no right-half-plane poles, if the passivity index $p(j\omega)$ in Table~\ref{T3:strength_metrics}  is positive for $\forall \omega \in[0,\infty)$, the system is stable, and $p(j\omega)$ can be regarded as a margin, as detailed in Ref.~\cite{xiongfei:passive}.  According to the inequality $\underline{\sigma}(A)\geq \underline{\lambda}(\frac{A+A^H}{2})$ in the Introduction, $p(j\omega)\leq \underline{\sigma}({{\bm Y}}_{\rm cl}(j\omega))$ is obtained. Thus the passivity margin provides a conservative lower bound on system strength.

\section{The Alignment of Device and System Level} 
{\subsection{Formulation of System Strength with an Added Device}}
Consider an additional converter connected to an interior bus $\left\{n+1\right\}$. This section attempts to evaluate the contribution of a converter connected at bus $\left\{n+1\right\}$ to the $n+1$-th bus strength as well as to the entire system strength. Assuming that the network lines have a uniform line constant $\tau$, a $2(n+1) \times 2(n+1)$ matrix according to~\eqref{eq12:additional} is,
  \vspace{-1mm}
\begin{equation}\label{eq17:Yn+1}
\begin{aligned}
{{\bm Y}}^{N+1}_{\rm Cl}(s)=\begin{bmatrix}
    {\bm Y}^{N}_{\rm de}(s) & {\bm 0}\\
    {\bm 0} & {Y}_{{\rm de},n+1}(s)
\end{bmatrix}+{{\bm B}}^{N+1}_{\rm Grid}\otimes {\gamma}(s)\,,
\end{aligned}
\end{equation}
where
$${{\bm B}}^{N+1}_{\rm Grid}=\begin{bmatrix}
{{\bm B}}^{N}_{\rm Grid}\in \mathbb{R}^{n \times n}  & {{\bm B}}^{N,1}_{\rm Grid}\in \mathbb{R}^{n \times 1}\\
  { {\bm B}}^{1,N}_{\rm Grid}\in \mathbb{R}^{1 \times n} & {{ B}^{N+1}_{\rm Grid}\in \mathbb{R}^{1 \times 1}}
\end{bmatrix}\,.$$

\begin{figure}
	\centering
	\includegraphics[width=2.5 in]{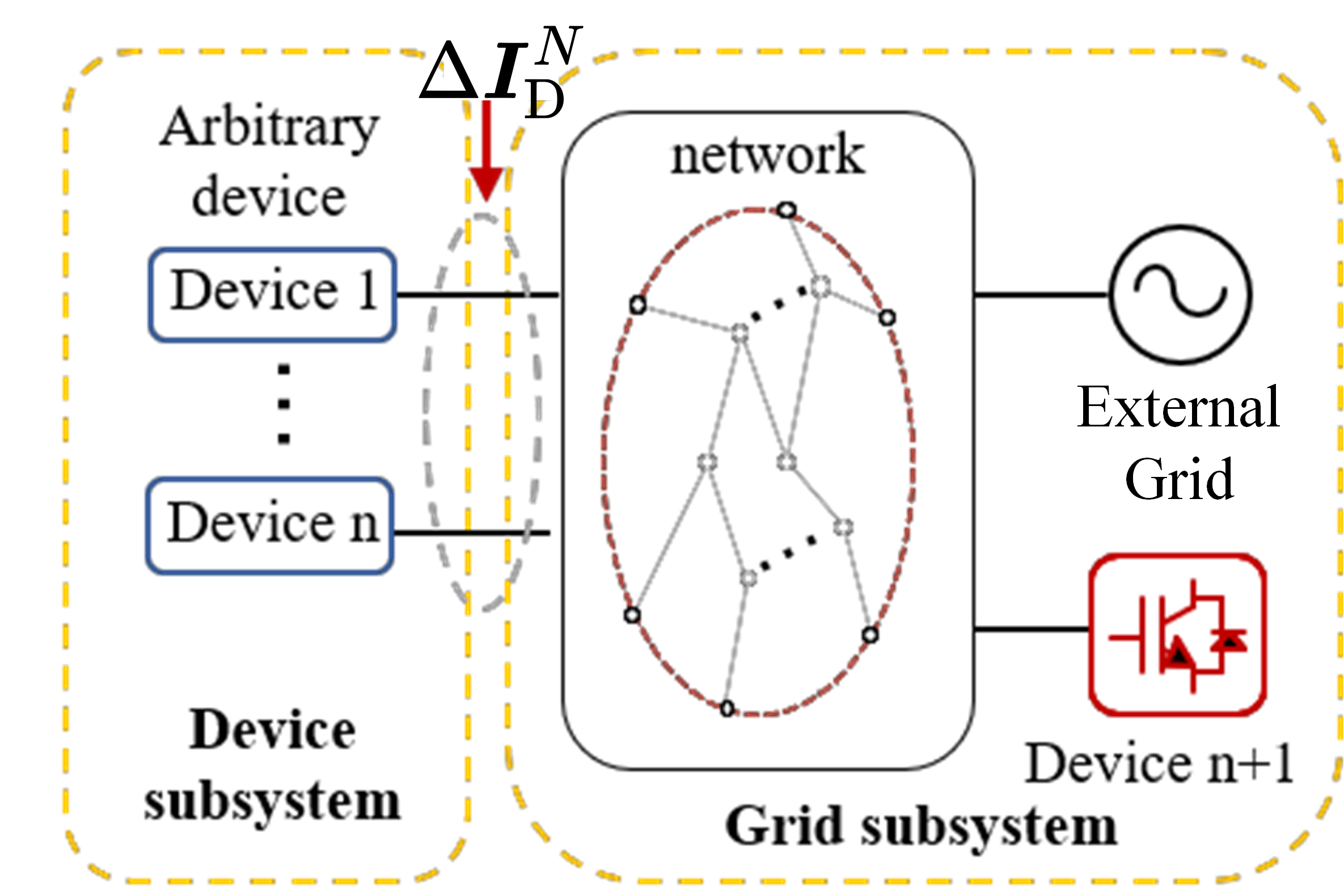}
	\vspace{-3mm}
	\caption{System structure diagram.} 
	\vspace{-0.4cm}
	\label{fig12:structure}
\end{figure}

\begin{lemma}[{System strength with an additional converter}] \label{adition strength}
Considering an additional converter connected to bus $\left\{n+1\right\}$ in Fig.~\ref{fig12:structure}, the system strength $\kappa(j\omega)$, grid strength $\alpha(j\omega)$, and bus strength $\kappa_{n+1}(j\omega)$ are respectively given by: 
\begin{equation}\label{eq:strength_addition}
\begin{aligned}
\kappa(j\omega)&=\underline{\sigma}\left[{{\bm Y}}^N_{\rm Cl}(j\omega)\right]\,,\\
\alpha(j\omega)& =\underline{\sigma}\left[ {{\bm Y}}^N _{\rm Grid}(j\omega)\left({I}_n \otimes {\gamma^{-1}_0}(j\omega)\right)\right]\,,\\
\kappa_{n+1}(j\omega)&=\left({\textstyle \sum_{j=1}^{n}\overline{\sigma}\left[{\bm Z}^{N,1}_{{\rm Cl},j}(j\omega)\right]+\underline{\sigma}^{-1} \left[{Y}^{N+1}_{\rm Cl}(j\omega)\right]}\right)^{-1}\,,
\end{aligned}
\end{equation}
where the explicit form is shown in Appendix~\ref{sec:appendixB}.
\end{lemma}

{ When extended to homogeneous multi-device integration, the same modeling approach can be adopted. Further details are omitted here for brevity.}

{\subsection{Linking FI to Bus Strength and System Strength}}
A GFM converter with $FI<1$ exhibits a stiff voltage characteristic, thereby enhancing the voltage stiffness of its connected bus. By eliminating the $n+1$-th bus via Kron reduction, the converter's admittance gets thus absorbed into the network admittance, as shown in Fig.~\ref{fig12:structure}, yielding ${\bm Y}^{N}_{\rm Grid}(s)$ as in~\eqref{eq27:Ygrideq}. The addition of a GFM converter enhances both the bus and the overall system strength.

\begin{proposition}[{GFM converter enhances bus strength, grid strength and system strength}]
\label{proof:align}
A GFM converter with $FI(j\omega)<1$ connected to bus $\{n+1\}$ enhances the grid strength $\alpha(j\omega)$ according to
\begin{equation}\label{eq:strength_enhance}
\alpha_{\rm new}(j\omega)\geq \alpha_{\rm old}(j\omega)+{\rm const.}\times(1-FI)\,,
\end{equation}
where $\alpha_{\rm new}(j\omega)$ and $\alpha_{\rm old}(j\omega)$ denote the grid strength with and without the additional converter at bus $\{n+1\}$, respectively. Consequently, this integration increases both the lower bound of the system strength $\kappa(j\omega)$ and that of the bus strength $\kappa_{n+1}(j\omega)$.

\end{proposition}

\begin{proof}
See Appendix~\ref{sec:appendixC}. 
\end{proof}
\vspace{-1mm}

{\begin{remark}
 In addition, according to the small-gain theorem~\cite{gainandphase}, system stability can be guaranteed if sufficient GFM converters are integrated to enhance grid strength, such that the grid strength exceeds the maximum gain of the device admittance $Y_{\rm de}(s)$. Moreover, in many cases, the integration of devices satisfying $FI<1$ improves not only a lower bound of system strength but also the system strength itself, as demonstrated by the case studies in Section V.
\end{remark}}

{\subsection{The application potential of the proposed metrics}

The proposed $FI$ and the system strength indices formalize previously loosely defined concepts regarding the GFM behavior and system strength. By establishing their interrelation, this paper bridges the device and system levels, thereby providing a theoretical foundation for both the design and placement of GFM converters. 

{\em 1) Control design of GFM converters}. $FI$ can serve as a principled cost function for GFM control design, which is fully compatible with the $\mathcal{H}_\infty$ robust control framework.

{Let $y$ be the controller input and $u$ be the controller output, and consider a control architecture given by $u=K(s)y$.} With the disturbance vector $\eta=\begin{bmatrix} U_{{\rm g}d} & U_{{\rm g}q}\end{bmatrix}$ and measured output vector $z=\begin{bmatrix} U_d & U_q\end{bmatrix}$, the plant model is given by

\begin{equation}\label{eq22:plant model}
\begin{bmatrix}
    z\\{u}
\end{bmatrix}=\begin{bmatrix} P_{11}(s) & P_{12}(s) \\ P_{21}(s) & P_{22}(s) \end{bmatrix} \begin{bmatrix}
    \eta\\{y}
\end{bmatrix}
\end{equation}

{$u=K(s)y$ and \eqref{eq22:plant model} can be used in a $\mathcal{H}_\infty$ control framework.} The closed-loop system dynamics can be expressed as 
\begin{equation}\label{eq23:closeofH}
\begin{aligned}
z & =\left\{P_{11}(s)+P_{12}(s) K\left[I-P_{22}(s) K\right]^{-1} P_{21}(s)\right\} \eta \\
& =:P(K)(s) \eta\,.
\end{aligned}
\end{equation}
Here $P(K)(s)$ denotes the sensitivity function under $\mathcal{H}_\infty$ control, which has $\overline{\sigma}[P(K)(j\omega)]=FI(j\omega)$. Selecting appropriate weighting functions $W(s)$ can optimize the $FI$ to achieve the GFM control design. 
\begin{equation}\label{eq24:Hcontrol}
\underset{\bm K}{\rm min} \,\,\underset{L_g\in \mathbb{L}}{\rm max}\,\,\|W(s)P(K)(s)\|_\infty\,.
\end{equation}
where $\mathbb{L}\subset \mathbb{R}$ denotes a wide range of SCRs. The structured controller $K(s)$ can be obtained via MATLAB’s \textit{hinfstruct}.

{\em 2)  Placement of GFM converters}. 
Future work can optimize the placement of GFM converters for maximizing system strength. A heuristic approach places GFM devices preferentially at weak buses with lower bus strength. A more principled approach formulates GFM placement via $\mathcal{H}_\infty$ norm optimization. Let the set of candidate buses for placing GFM converters be $\mathcal{V} =\left\{n+1,\cdots,n+m\right\}$. The diagonal capacity matrix of buses at $\mathcal{V}$ is  ${\bm S}_{{\rm B},M}$, and the system strength is $\kappa(j\omega)=\overline{\sigma}^{-1}[{\bm Y}^{-1}_{\rm Cl}({\bm S}_{{\rm B},M})(j\omega)]$. Then, the GFM placement optimization problem can be formulated as: 
\begin{equation}\label{eq25:placement}
\begin{aligned}
    \underset{{\bm S}_{{\rm B},M}}{\rm min}\,\|{\bm Y}^{-1}_{\rm Cl}({\bm S}_{{\rm B},M})(s)\|_\infty\\
    {\rm subject \,\,to}\,  \sum ^{m}_{i=1}  S_i\leq S_{\rm tot},
\end{aligned}
\end{equation}
where $ S_{\rm tot}$ is the total capacity limit of GFM  converters.

Moreover, if information on other devices is unavailable or the optimization encounters dimensionality challenges, grid strength can be used as an alternative cost function to simplify the optimization.

 \begin{figure} 
	\centering
	\includegraphics[width=2.5 in]{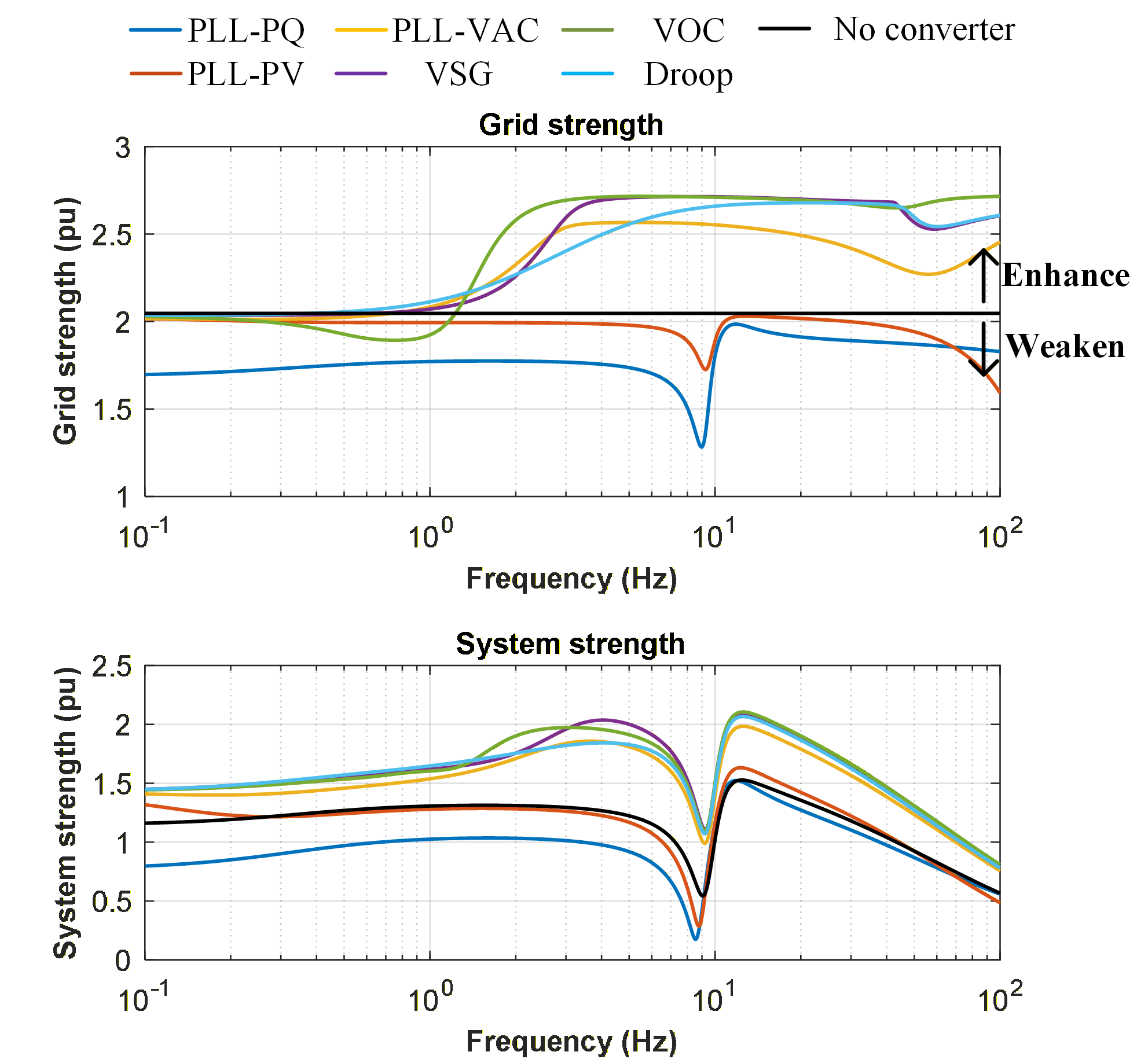}
	\vspace{-3mm}
	\caption{Different control of converter at bus 38 in the IEEE 39-bus system: grid strength and system strength.} 
	\vspace{-0.4cm}
	\label{fig13:strength}
\end{figure}
 \begin{figure} 
	\centering
	\includegraphics[width=2.6 in]{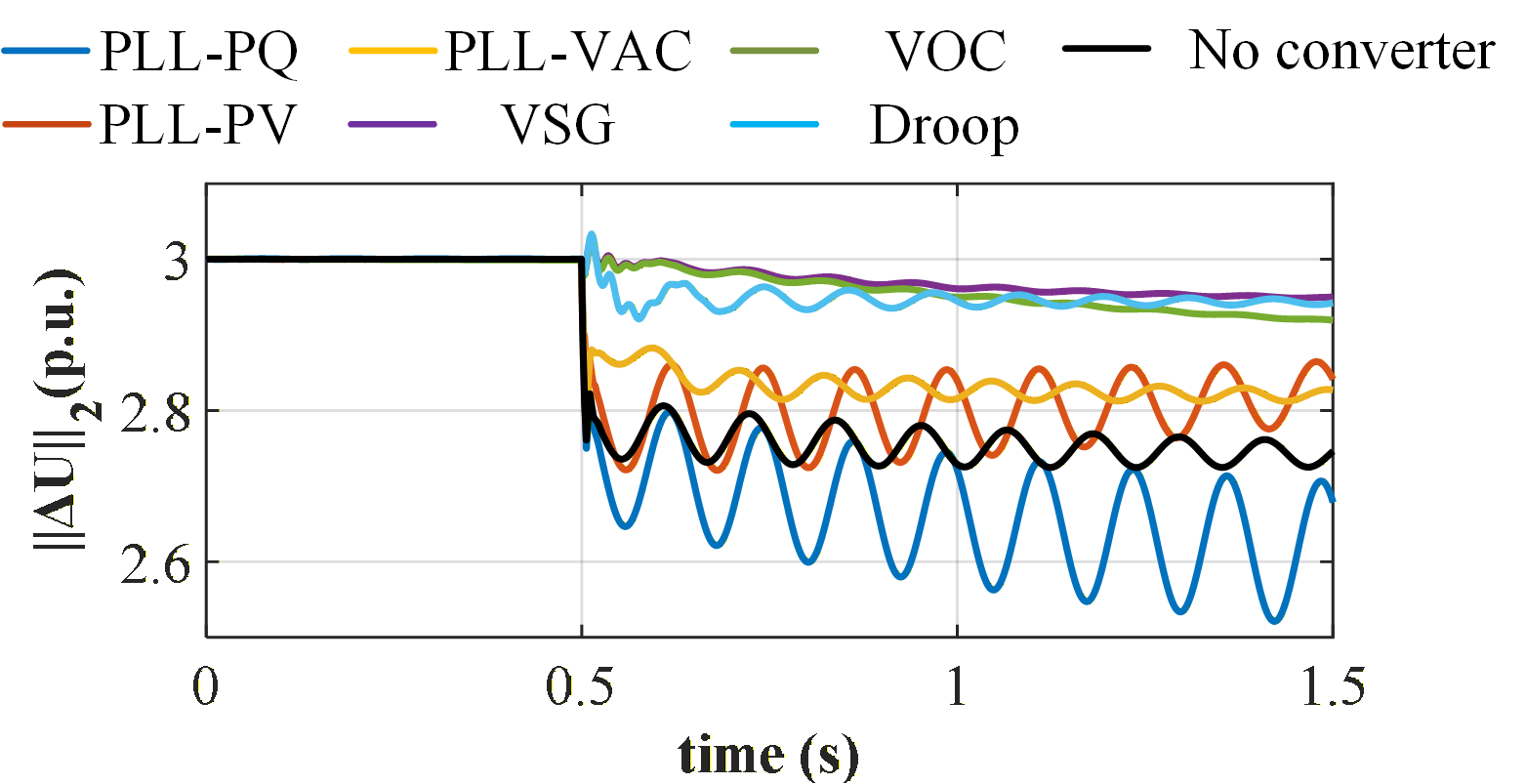}
	\vspace{-3mm}
	\caption{Time-domain response waveforms.} 
	\vspace{-0.4cm}
	\label{fig14:time}
\end{figure}

{\em 3)  Prospects for 1D-VS behavior}. This paper focuses on the 2D-VS behavior to distinguish GFM from GFL. Further characterization of GFL converters is needed to capture their 1D-VS behavior in either voltage or frequency, enabling a more detailed classification of GFL devices into voltage-supporting converters (1D-VS in voltage), frequency-supporting converters (1D-VS in frequency), and fully GFL converters. Similarly, system strength requires further refinement to quantify system voltage strength and system frequency strength. These directions represent important avenues for future research.

{\em 4)  The measurement of $FI$.} The proposed $FI$ can also be evaluated for arbitrary devices, rather than only those considered in this paper, through measurement-based admittance identification. By measuring the equivalent grid impedance and device admittance, $FI$ can be calculated without requiring detailed internal models of the converter.}

{\em 5)  Extension to Other Scenarios.} The proposed concepts of $FI$ and system strength are not restricted to the specific devices and operating scenarios considered in this paper. Both metrics are formulated based on singular-value characterizations of the underlying input--output behavior, and their assessment is therefore not tied to the particular analytical tools adopted here. In addition to passivity, gSCR, and ESCR, alternative frameworks, such as small-gain, small-phase, and SRG analyses~\cite{gainandphase}, can provide complementary bounds or certificates for the same underlying input--output properties and can thus be related to the proposed $FI$/strength framework. This perspective also suggests a natural extension beyond the small-signal regime. Under large disturbances, nonlinear input--output gain bounds, for example those obtained from suitably constructed storage functions, may provide a basis for extending $FI$ and system strength quantification to the large-signal setting. Such extensions are left for future investigation.

\section{Case Results}

{{\subsection{Case studies on the 39-bus system}}}
    \begin{figure*} 
	\centering
	\includegraphics[width=5.9in]{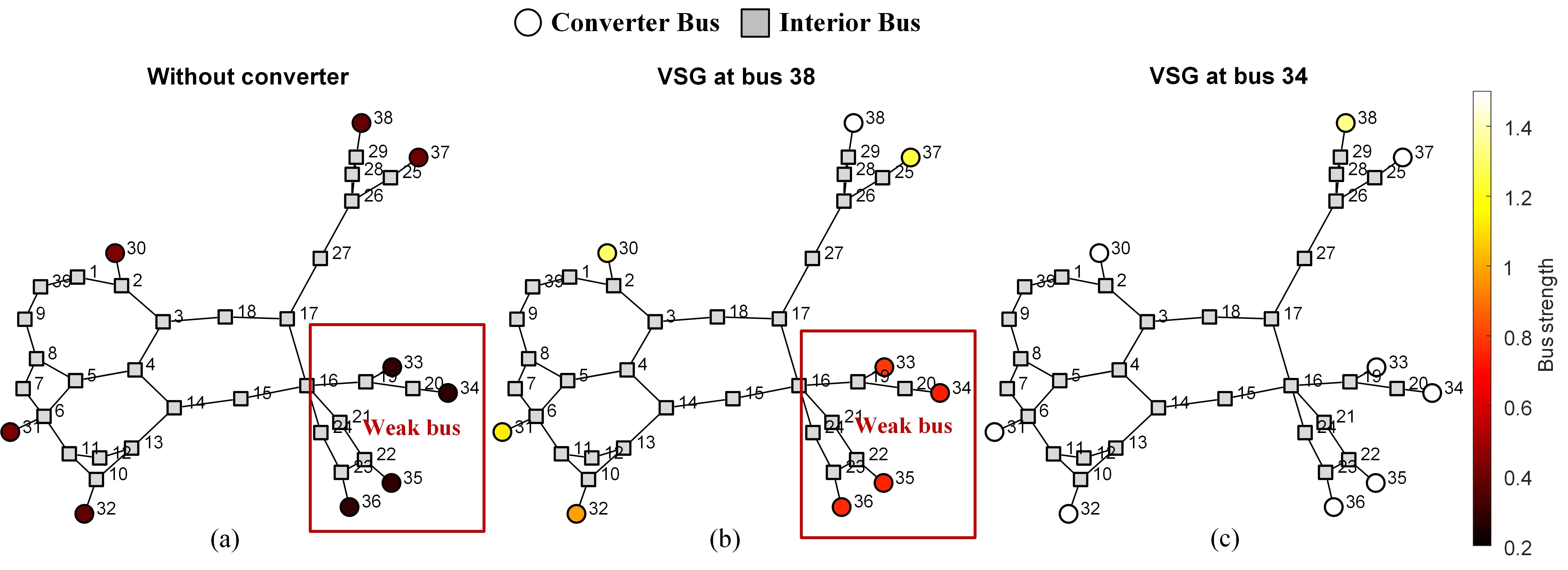}
	\vspace{-3mm}
	\caption{The bus strength of the IEEE 39-bus system. (a) without additional converter. (b) a VSG connected at bus 38. (c) a VSG connected at bus 34.} 
	\vspace{-0.4cm}
	\label{fig15:busstrength}
\end{figure*}
In order to demonstrate the utility of the proposed indices, the IEEE 39-bus system in \cite{kehao:dual_axis} is used for validation. PLL-PQ converters are connected to buses $\left\{30\sim37\right\}$ and treat bus $\left\{39\right\}$ as an infinite bus. An additional converter with different control strategies in Fig.~\ref{fig2:control} is connected to bus $\left\{38\right\}$ to check the variation in system strength, grid strength and bus strength.

The variations in system strength and grid strength, both with and without converter connections at bus $\left\{38\right\}$, are shown in Fig. \ref{fig13:strength}. It is evident that grid strength and system strength are improved by enhancing (i.e., lowering) the $FI$ of that bus, consistent with the $FI$ discussed in Proposition \ref{proof:align}. The GFM converters (VSG, VOC, droop, and PLL-VAC) all contribute to enhancing system strength, whereas the GFL converters (PLL-PQ and PLL-PV) reduce system strength.  A smaller $FI$ corresponds to greater grid strength, and thus also to an improvement in system strength. 

 \begin{figure} 
	\centering
	\includegraphics[width=3 in]{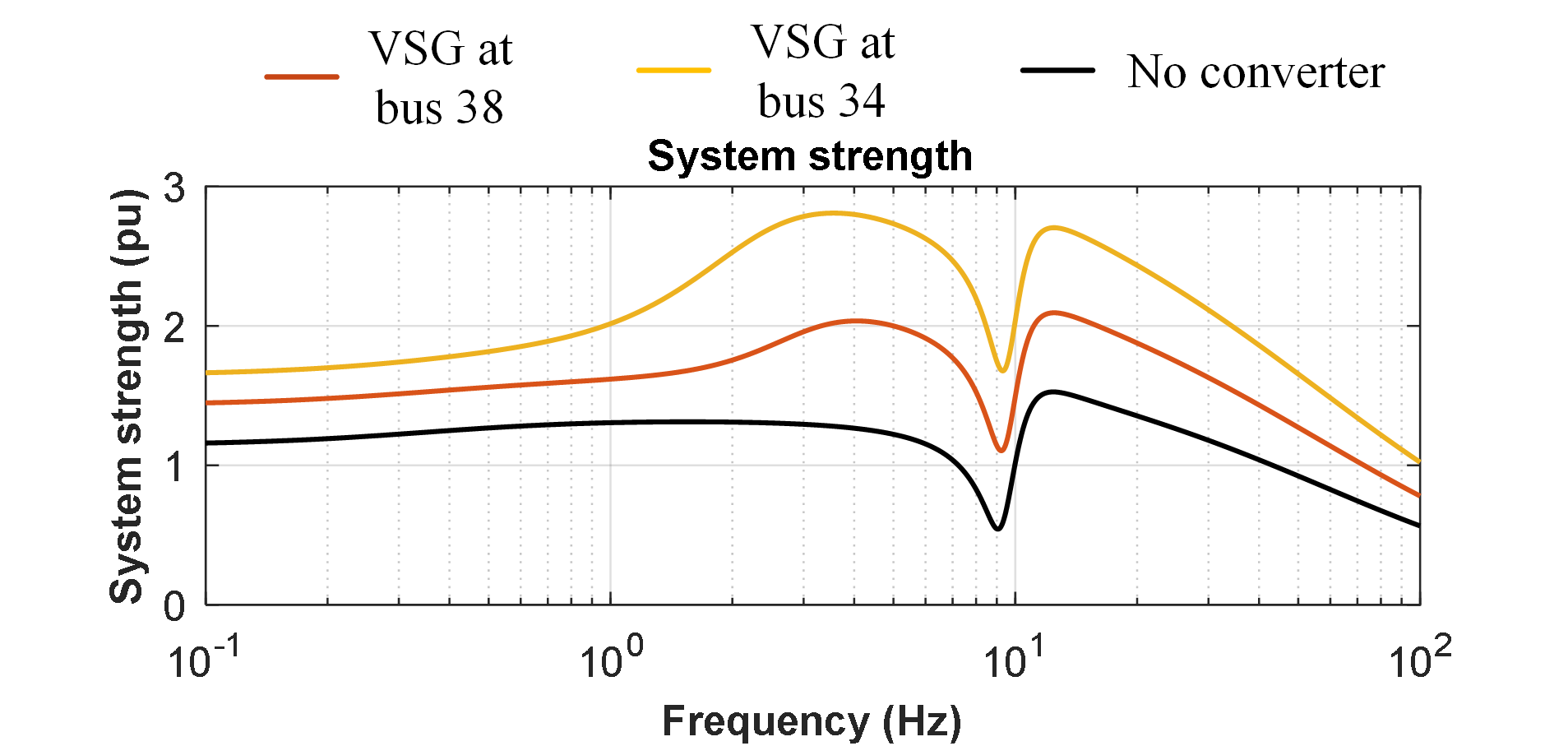}
	\vspace{-3mm}
	\caption{VSG connected at different buses in the IEEE 39-bus system: grid strength and system strength.} 
	\vspace{-0.4cm}
	\label{fig16:sstrength3438}
\end{figure}
 \begin{figure} 
	\centering
	\includegraphics[width=2.8 in]{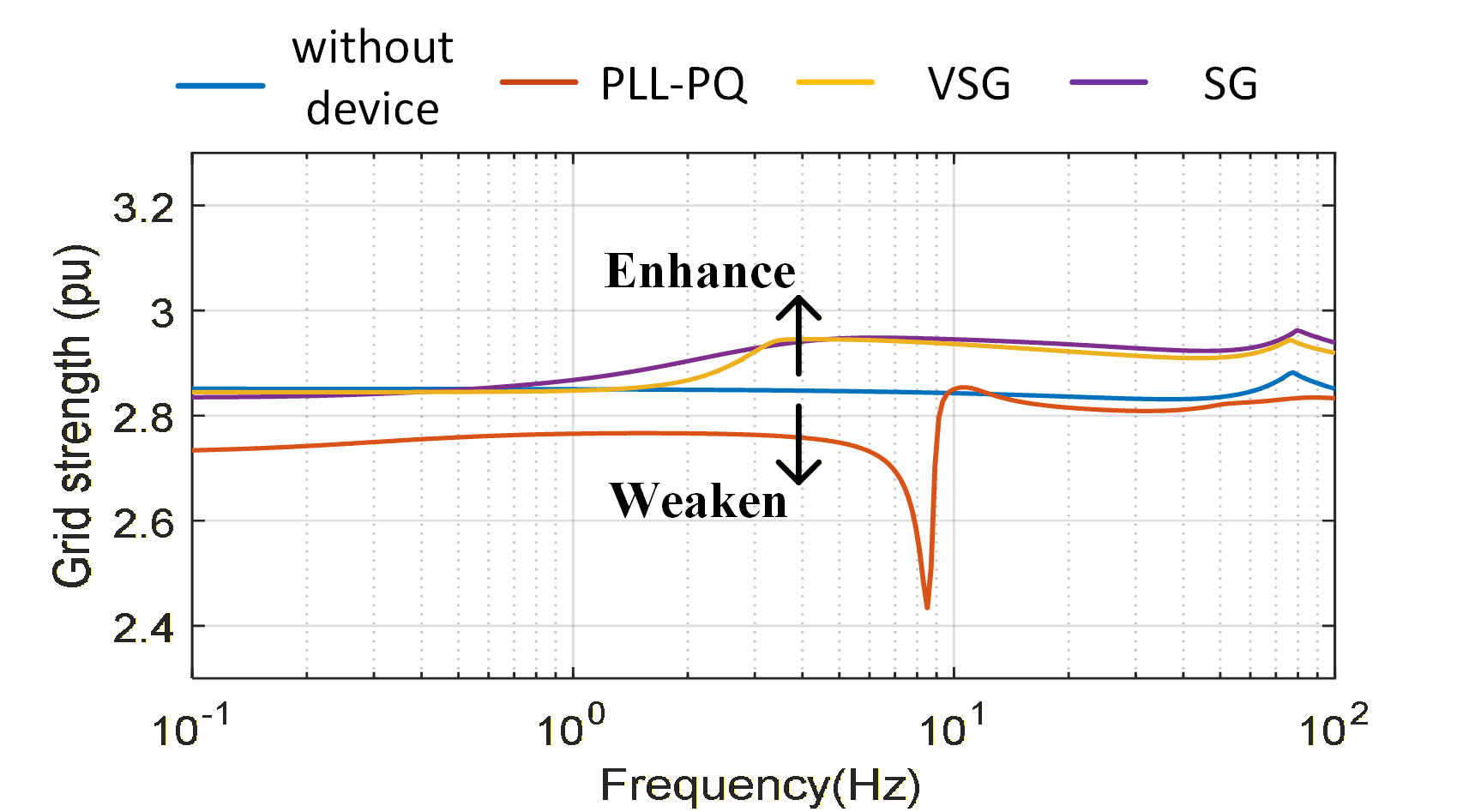}
	\vspace{-3mm}
	\caption{{Different devices at bus 12 in the IEEE 68-bus system: grid strength.}} 
	\vspace{-0.4cm}
	\label{fig16:68strength}
\end{figure}

In addition, a step current disturbance with an amplitude of 1 p.u., is injected at bus $\left\{9\right\}$ at $t=0.5\rm s$ . The time-domain response of the norm of the voltage vector of buses $\left\{30,\cdots,38\right\}$ is shown in Fig.\ref{fig14:time}. The time-domain simulations further demonstrate that the higher system strength leads to smaller voltage oscillations, indicating better disturbance rejection, consistent with our previous theory findings.

Figs. \ref{fig15:busstrength} (a), and (b) present the bus strength at 9Hz (the frequency with the lowest system strength) under two scenarios: without any converter connected at bus $\left\{38\right\}$, and with a VSG connected at bus $\left\{38\right\}$. It can be seen that connecting a VSG to bus $\left\{38\right\}$, significantly increases its bus strength  (approximately 4.57 pu at 9 Hz), and also improves the bus strengths across all other buses. 

In addition, the results indicate that the area surrounding buses $\left\{33 \sim 36\right\}$ is the weakest region, limiting the overall system strength. Therefore, the VSG initially connected at bus $\left\{38\right\}$ is swapped with the PLL-PQ at bus $\left\{34\right\}$. The resulting bus strengths at 9 Hz after the swap are shown in Fig. \ref{fig15:busstrength} (c), and a comparison of system strength before and after the swap is illustrated in Fig. \ref{fig16:sstrength3438}. This demonstrates that enhancing system strength by connecting a GFM converter to the weakest bus is more effective than placing it at other locations. 

Furthermore, the weakest bus strength is slightly lower than the overall system strength. This confirms that if the bus strength at each individual bus meets a required threshold (e.g., not less than 0.5, and preferably greater than 1), the overall system strength can be guaranteed to meet the standard. 

\begin{figure*} 
	\centering
	\includegraphics[width=6.2in]{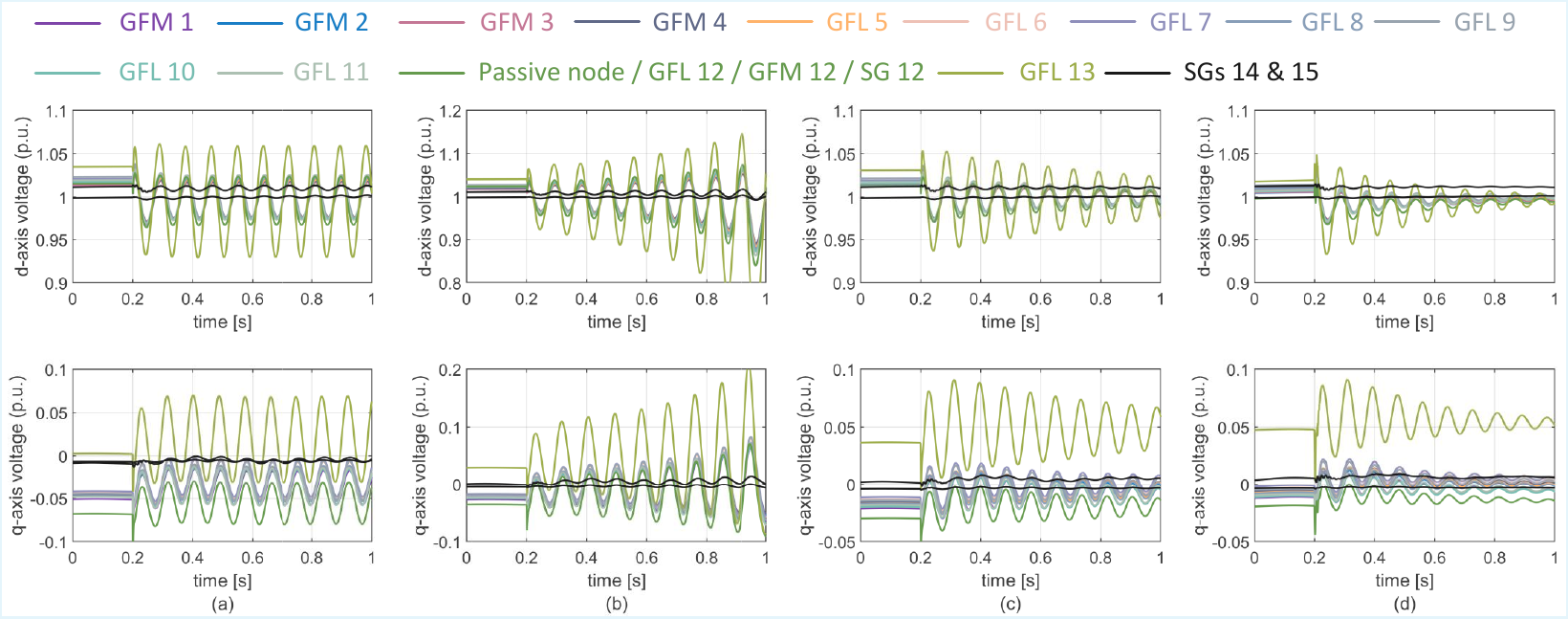}
	\vspace{-3mm}
	\caption{{The voltage responses in global dq frame at bus $\{1\sim15\}$ of the 68-bus system. (a) without additional converter; (b) a PLL-PQ converter connected; (c) a VSG connected; (d) an SG connected.} }
	\vspace{-0.4cm}
	\label{fig16:68bustime}
\end{figure*}
{{\subsection{Case studies on the 68-bus system}}}

{The proposed method is further validated on the 68-bus test system (refer to Fig.12 in~\cite{gainandphase}). The system consists of both converters and synchronous generators. Specifically, buses $\left\{1\sim4\right\}$ are equipped with VSG, buses $\left\{5\sim13\right\}$ are connected with PLL-PQ converters, buses $\left\{14,15\right\}$ are SGs, and bus $\left\{16\right\}$ is modeled as an external grid. The network model includes shunt capacitors to provide a more realistic representation of the grid.}

{To evaluate the impact of different devices on grid strength, bus 12 is selected as the test location. Four scenarios are considered: 1) bus $\left\{12\right\}$ is left as a node without any connected devices; 2) a PLL-PQ converter is connected at bus $\left\{12\right\}$; 3) a VSG-based converter with $L_g \approx 0.12$pu is connected at bus $\left\{12\right\}$; 4) an SG is connected at bus $\left\{12\right\}$.}

{According to the previous analysis, there is $FI_{\mathrm{SG}} < FI_{\mathrm{VSG}}(L_g \approx 0.12{\rm pu}) < 1 < FI_{\mathrm{PLL\text{-}PQ}}$ over the frequency range of interest. This indicates that the expected contribution to grid strength follows the order: SG, VSG, without device, PLL-PQ converter. The grid strength under different scenarios is calculated and shown in Fig.~\ref{fig16:68strength}. It can be observed that the SG connection provides the largest grid strength enhancement, while the improvement achieved by the VSG is smaller than that of the SG. In contrast, the connection of the PLL-PQ converter reduces the grid strength. These results are consistent with the theoretical analysis.}

{To verify this, time-domain simulations are conducted. At $t = 0.2$ s, a disturbance is introduced by applying a step change to the grid-side impedance of the converter. The voltage responses in the global dq frame under the four scenarios are then compared in Fig.~\ref{fig16:68bustime}. When bus $\left\{12\right\}$ is left as a node without device, the system exhibits sustained oscillations with nearly constant amplitude, indicating marginal stability. When a PLL-PQ converter is connected at bus $\left\{12\right\}$, the oscillations become divergent, and the system loses stability. In contrast, when either a VSG or an SG is connected at bus $\left\{12\right\}$, the oscillations are damped and the voltage response converges. Moreover, the SG case shows a smaller oscillation amplitude and stronger damping than the VSG case. Overall, this case study demonstrates that the proposed analysis can effectively predict the impact of different device types on grid strength and time-domain stability.}

\section{Conclusions}
At the device-level, the proposed $FI$ quantifies the GFM behavior by characterizing 2D-VS behavior of a converter across different frequencies. This paper also demonstrates that GFM behavior can be achieved using a PLL-based structure. At the system-level, the system strength, defined as the gain from multi-bus current or power disturbances to multi-bus voltage output, describes 2D-VS characteristics of the system's multiple buses, which essentially characterizes the stability, disturbance rejection, and robustness of power systems. The proposed grid strength and bus strength can effectively indicate the weakest bus, and ensuring sufficient strength at each bus is sufficient to meet the overall system strength requirements. The results demonstrate that a converter exhibiting GFM behavior at the device level also enhances system strength. In words, a device exhibiting 2D-VS behavior will enhance the 2D-VS characteristics of the system's multiple buses. The proposed metrics can be used to guide control design and placement of GFM converters, and the real-time monitoring of system stability. 

Future research should also investigate the detailed decoupling of voltage and frequency, investigating both the device voltage/frequency 1D-VS behavior and the system voltage/frequency strength. {  Moreover, the proposed indices are currently intended for small-signal operation around normal operating points and may have limitations under nonlinear conditions such as FRT operation or current limiting. Therefore, extending the framework to quantify large-signal GFM capability and system strength under nonlinear operating conditions remains an important direction for future work.}

{\appendices

{ 
\section{{Models and Parameters}} \label{sec:model}

\subsection{SG and converter dynamics} \label{sec:SG1}}

{ 1) SG dynamics.}

{ The SG model, together with its turbine--governor, AVR--excitation system, and PSS dynamics, is illustrated in Fig.~\ref{fig19:sgmodel}. The winding dynamics, and the power calculation are formulated as follows
\begin{equation}
\begin{array}{rcl}
{\rm (Winding)} &&
\left\{
\begin{aligned}
&T_D'\dot{E}_Q' = -a_DE_Q'+(a_D-1)E_Q''+E_{f},\\
&T_D''\dot{E}_Q'' = E_Q'-E_Q''-(X_D'-X_D'')I_D,\\
&T_Q'\dot{E}_D' = -a_QE_D'+(a_Q-1)E_D'',\\
&T_Q''\dot{E}_D'' = E_D'-E_D''+(X_Q'-X_Q'')I_Q.\\
&\varphi_D = E_Q''-X_D''I_D,\, \varphi_Q=-E_D''-X_Q''I_Q,\\
&U_D = -\varphi_Q-R_aI_D,\\
&U_Q=\varphi_D-R_aI_Q.
\end{aligned}\right.
\\
\\
\text{(Power Cal.)} &&
\left\{
\begin{aligned}
P &= U_D I_D+U_Q I_Q,\\
Q &= -U_D I_Q+U_Q I_D,
\end{aligned}
\right.
\end{array}
\label{eq:sg_electromagnetic}
\end{equation}
where $T_D'$ ($T_Q'$) and $T_D''$ ($T_Q''$) are the $D$-axis ($Q$-axis) transient and subtransient time constants, $E_D'$ ($E_Q'$) and $E_D''$ ($E_Q''$) are the $D$-axis ($Q$-axis) transient and subtransient internal voltages, $X_D'$ ($X_Q'$) and $X_D''$ ($X_Q''$) are the $D$-axis ($Q$-axis) transient and subtransient reactances, $R_a$ is the stator resistance, $I_D$ and $I_Q$ are the $D$-axis and $Q$-axis stator currents, $U_D$ and $U_Q$ are the $D$-axis and $Q$-axis stator voltages, $\varphi_D$ and $\varphi_Q$ are the $D$-axis and $Q$-axis stator fluxes, $E_{f}$ is the excitation voltage, $a_D=(X_D-X_D'')/(X_D'-X_D'')$, and $a_Q=(X_Q-X_Q'')/(X_Q'-X_Q'')$.}

{ Fig.~\ref{fig19:sgmodel} and~\eqref{eq:sg_electromagnetic} together constitute the dynamic model of an SG, which is linearized and subsequently used to calculate the proposed $FI$ and strength.}
 \begin{figure} 
	\centering
     
	\includegraphics[width=2.8 in]{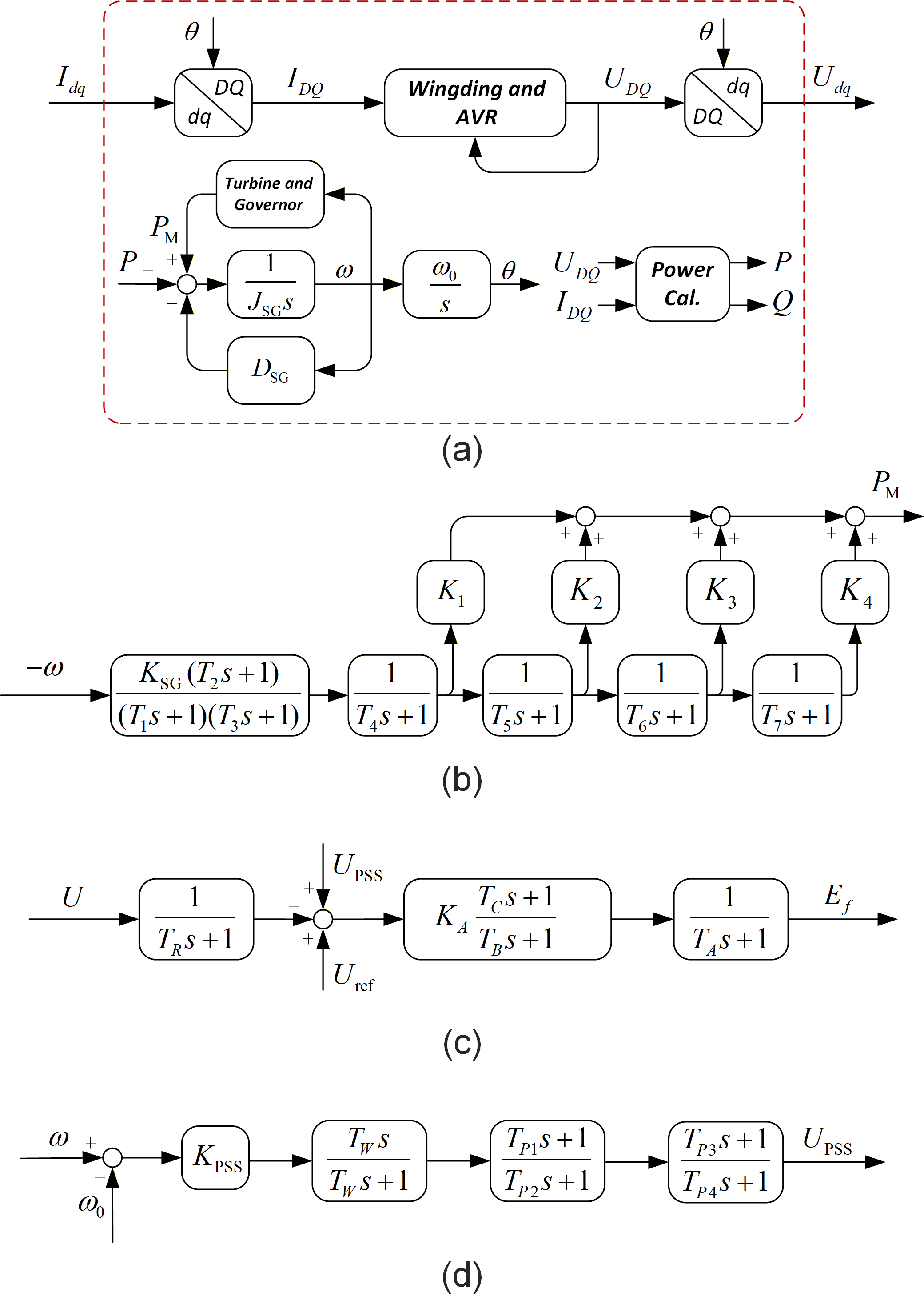}
	\vspace{-3mm}
	\caption{{The equivalent block of SG model. (a) Model of an SG. (b) turbine--governor. (c) AVR--excitation. (d) PSS.}} 
	\vspace{-0.4cm}
	\label{fig19:sgmodel}
\end{figure}

{ 2) Converter dynamics.}

{ The converter dynamics considered in the analysis and numerical calculations of this paper are shown in Figs.~\ref{fig1:converter} and~\ref{fig2:control}, including detailed filter dynamics, current control, synchronization control, voltage control, and power control. The virtual impedance $Z_v(s)$ and virtual admittance $Y_v(s)$ in Fig.~\ref{fig2:control} (b) and~(c), respectively, are given by
\begin{equation}\label{eq:zvyv}
\begin{aligned}
Z_v(s)=\frac{1}{1+T_zs}\begin{bmatrix} R_v & -L_v \\
 L_v & R_v \end{bmatrix}
\,,\,
Y_v(s)=\frac{1}{1+T_ys}\begin{bmatrix} 0 & y_v \\
 y_v & 0 \end{bmatrix}.
\end{aligned}
\end{equation}}

{ 
\subsection{The small-signal admittance model of SG and converters}} \label{sec:dqmodel}
{ This subsection describes how the device small-signal admittance models used for $FI$ and strength calculations are obtained. Considering the SG and converter dynamics introduced in Appendix~\ref{sec:SG1} and illustrated in Figs.~\ref{fig1:converter},~\ref{fig2:control}, and~\ref{fig19:sgmodel}, the terminal voltage vector $[U_d\,\,U_q]^{\mathsf T}$ is selected as the input and the terminal current vector $[I_d\,\,I_q]^{\mathsf T}$ as the output. The corresponding nonlinear state-space models derived from Figs.~\ref{fig1:converter}, \ref{fig2:control}, and \ref{fig19:sgmodel} are then linearized around the operating point using a first-order Taylor expansion, yielding
\begin{equation}\label{eq:state}
\begin{aligned}
\dot{\bm x} = {\bm A}{\bm x}+{\bm B}\begin{bmatrix}
    \Delta U_d\\ \Delta U_q
\end{bmatrix},\,\,\begin{bmatrix}
    \Delta I_d\\ \Delta I_q
\end{bmatrix}={\bm C}{\bm x}+{\bm D}\begin{bmatrix}
    \Delta U_d\\ \Delta U_q
\end{bmatrix}.
\end{aligned}
\end{equation}}

{ Consequently, the small-signal admittance matrix used in the $FI$ and strength calculations can be given by
\begin{equation}\label{eq:state}
\begin{aligned}
Y_{\rm de}(s)={\bm C}(s{\bm I}-{\bm A})^{-1}{\bm B}+{\bm D}.
\end{aligned}
\end{equation}}

{ The explicit admittance expressions for different types of devices can be found in~\cite{WenBo:dq_model} (PLL-PQ),~\cite{7:howmany} (PLL-PV, VSG, droop),~\cite{dvoc_dq} (dVOC), and~\cite{sgimpedance} (SG).}

{ 
\subsection{The parameters of devices}} \label{sec:parameters}

{ See Table~\ref{tab:single_converter_parameters}.}

\begin{table}
\centering
           
\caption{Parameters of converters and SG}
\label{tab:single_converter_parameters}
\footnotesize

\renewcommand{\arraystretch}{1.38}
\setlength{\tabcolsep}{3.8pt}

\resizebox{0.98\columnwidth}{!}{%
\begin{tabular}{|@{\;}ll@{\qquad}ll@{\;}|}
\hline
\multicolumn{4}{|c|}{\bfseries Base Values for Per-unit Calculation}
\\ \hline

$f_{\mathrm{base}}$ & $=50\,\mathrm{Hz}$ & $\omega_{\mathrm{base}}$ & $=2\pi f_{\mathrm{base}}$
\\
$U_{\mathrm{base}}$ & $=380\,\mathrm{V}$ & $S_{\mathrm{base}}$ & $=1\,\mathrm{MVA}$
\\ \hline

\multicolumn{4}{|c|}{\bfseries Common Parameters of the Converters (Per-unit Values)}
\\ \hline

\multicolumn{2}{|l}{Converter-side inductors: $L_f=0.05$} &
\multicolumn{2}{l|}{LCL capacitors: $C_f=0.05$}
\\

\multicolumn{2}{|l}{PI gains of the current loop: $1.5,\,10$}&
\multicolumn{2}{l|}{VF time constant: $T_{v}=0.001$}
\\ \hline

\multicolumn{4}{|c|}{\bfseries The Parameters of the PLL-PQ and PLL-PV (Per-unit Values)}
\\ \hline

\multicolumn{2}{|l}{PLL PI gains: $10,\, 3500$} &
\multicolumn{2}{l|}{PI gains of outer loop: $2,\, 4$}
\\ \hline

\multicolumn{4}{|c|}{\bfseries The Parameters of the VSG and droop (Per-unit Values)}
\\ \hline

\multicolumn{2}{|l}{Inertia and damping of VSG: $2,\,50$} &
\multicolumn{2}{l|}{droop coefficient: $K_P=1/50$}
\\ \hline

\multicolumn{2}{|l}{PI gains of DQ voltage control: $0.5,\,2$} &
\multicolumn{2}{l|}{Virtual impedance: $L_v=0.1$}
\\ \hline

\multicolumn{2}{|l}{Virtual impedance: $R_v=0$} &
\multicolumn{2}{l|}{Virtual impedance: $T_z=0.001$}
\\ \hline

\multicolumn{4}{|c|}{\bfseries The Parameters of the dVOC (Per-unit Values)}
\\ \hline

\multicolumn{2}{|l}{Reactive droop: $K_Q=0.02$} &
\multicolumn{2}{l|}{Voltage droop: $K_U=5$}
\\ \hline

\multicolumn{2}{|l}{Active droop: $K_P=0.02$} &
\multicolumn{2}{l|}{Voltage setting: $U_0=1$}
\\ \hline

\multicolumn{4}{|c|}{\bfseries The Parameters of the PLL-VAC (Per-unit Values)}
\\ \hline

\multicolumn{2}{|l}{Inertia and damping: $2,\,50$} &
\multicolumn{2}{l|}{PI gains of power loop: $2,\, 4$}
\\ \hline

\multicolumn{2}{|l}{Virtual admittance: $y_v=10$} &
\multicolumn{2}{l|}{Virtual admittance: $T_y=0.001$}
\\ \hline

\multicolumn{4}{|c|}{\bfseries The Parameters of the SG (Per-unit Values)}
\\ \hline

\multicolumn{2}{|l}{Inertia and damping: $8,\,0.1$} &
\multicolumn{2}{l|}{Voltage setting: $U_{\rm ref}=1.01$}
\\ \hline

\multicolumn{4}{|l|}{Turbine-governor $T_1-T_7$: $0.2,\,0,\,0.1,\,0.2,\,6,\,6,\,0.4$} 
\\ \hline

\multicolumn{4}{|l|}{Turbine-governor $K_{\rm SG},\,K_1-K_4$: $20,\,0.22,\,0.22,\,0.3,\,0.26$} 
\\ \hline

\multicolumn{4}{|l|}{AVR--excitation $K_A,\,T_A-T_C,\,T_R$: $200,\,0.015,\,10,\,1,\,0.01$} 
\\ \hline

\multicolumn{4}{|l|}{PSS $K_{\rm PSS},\,T_{P1}-T_{P4},\,T_W$: $20,\,1,\,0.2,\,1,\,0.1,\,10$} 
\\ \hline

\multicolumn{4}{|l|}{ $D$-axis impedance $R_a,\,X_D,\,X'_D,\,X''_D$: $0.003,\,0.362,\,0.06,\,0.048$} 
\\ \hline

\multicolumn{4}{|l|}{ $Q$-axis impedance $R_a,\,X_Q,\,X'_Q,\,X''_Q$: $0.003,\,0.352,\,0.13,\,0.048$} 
\\ \hline

\multicolumn{4}{|l|}{(Sub)transient time constant $T'_D,\,T'_Q,\,T''_D,\,T''_Q$: $8,\,1,\,0.03,\,0.07$} 
\\ \hline

\end{tabular}%
}
\end{table}

\section{{Proof of Lemma.~\ref{FIandpassivity}}}
\label{sec:appendixA}
{$FI(j\omega)<1$ is equivalent to $\underline{\sigma}\left(I_2+L_gH(j\omega)\right)>1$, i.e., 
\[
\begin{aligned}
0<&\|\left(I_2+L_gH(j\omega)\right)x\|^2_2-\|x\|^2_2\\
=&x^H\left(I_2+L_gH(j\omega)\right)^H\left(I_2+L_gH(j\omega)\right)x-x^Hx\\
=&x^H(L_gH^H(j\omega)+L_gH(j\omega))x+x^HL^2_gH^H(j\omega)H(j\omega)x\\
=&2L_g\mathfrak{R} \left(x^HH(j\omega)x\right)+L^2_g\|H(j\omega)x\|^2_2\,,
\end{aligned}   
\]
for $\forall x \neq 0,\ x \in \mathbb{C}^2$ and $\forall L_g \in (0,1)$. After canceling $L_g$ from  this inequality, it holds true if $H(j\omega)+H^H(j\omega)\succ0$. This condition is also necessary for $FI(j\omega)<1$ in the limit of $L_g \to0$.

}

\section{The Expression of Eq.~\eqref{eq:strength_addition}}
\label{sec:appendixB}
The inverse of a block matrix $A=\left[\begin{smallmatrix}  A_1 & A_2\\A_3 & A_4 \end{smallmatrix}\right]$ is given by,
  \vspace{-1mm}
\begin{equation}\label{eq26:inverse}
\begin{aligned}
A^{-1}=\begin{bmatrix} (A_1-T_1)^{-1} & -A^{-1}_1A_2(A_4-T_2)^{-1} \\
 -(A_4-T_2)^{-1}A_3A^{-1}_1 & (A_4-T_2)^{-1} \end{bmatrix}
\,,
\end{aligned}
\end{equation}
where $T_1=A_2A^{-1}_4A_3$, and  $T_2=A_3A^{-1}_1A_2$, and provided that all inverses exist.

Let $A={\bm Y}^{N+1}_{\rm Cl}(s)$ in~\eqref{eq17:Yn+1}, thus we obtain ${{\bm Z}}^{N+1}_{\rm Cl}(s)=\left[{\bm Y}^{N+1}_{\rm Cl}(s)\right]^{-1}=$ 
  \vspace{-1mm}

\begin{equation}\label{eq18:Zn+1}
\begin{aligned}
\begin{bmatrix}
\left ({\bm Y}^{N}_{\rm de}(s)+{{\bm Y}}^{N}_{\rm Grid}(s)\right)^{-1} &{{\bm Z}}^{N,1}_{\rm Cl}(s)\\
   {{\bm Z}}^{1,N}_{\rm Cl}(s) &  {\left({Y}_{{\rm de},n+1}(s)+{{ Y}}^{N+1}_{\rm Grid}(s)\right)}^{-1 }
\end{bmatrix}\,,
\end{aligned}
\end{equation}

with
  \vspace{-1mm}
\begin{equation}\label{eq27:Ygrideq}
\begin{aligned}
{\bm Y}^{N}_{\rm Grid}(s)&:={\bm B}^{N}_{\rm Grid}\otimes \gamma(s)-T_1\\
&={\bm B}^{N}_{\rm Grid}\otimes \gamma(s)-{\bm B}_{eq} \otimes {S}_v(s)\gamma(s)\\
{\bm Y}^{N}_{\rm Cl}(s)&:={\bm Y}^{N}_{\rm de}(s)+{{\bm Y}}^{N}_{\rm Grid}(s)\\
{Y}^{N+1}_{\rm Grid}(s)&:=
{ B}^{N+1}_{\rm Grid}\otimes {\gamma}(s)-T_2\\
{Y}^{N+1}_{\rm Cl}(s)&:={Y}_{{\rm de},n+1}(s)+{{ Y}}^{N+1}_{\rm Grid}(s)\\
&=\left[{ B}^{N+1}_{\rm Grid}\otimes {\gamma}(s)\right]S^{-1}_v(s)-T_2
\\
{\bm Z}^{N,1}_{\rm Cl}(s)&:=-A^{-1}_1A_2\left[{Y}^{N+1}_{\rm Cl}(s)\right]^{-1}\\
{\bm Z}^{1,N}_{\rm Cl}(s)&:=-\left[{Y}^{N+1}_{\rm Cl}(s)\right]^{-1}A_3A^{-1}_1\,,
\end{aligned}
\end{equation}
where ${\bm B}_{eq}=\left({{\bm B}^{N,1}_{\rm Grid}{\bm B}^{1,N}_{\rm Grid}}\right)/{{B}^{N+1}_{\rm Grid}}$, $S_v(s)$ is identical to the expression in~\eqref{eq3:sensitivity of single VSC} with $L_{\rm g}=1/\tilde{ B}^{N+1}_{\rm Grid}$ and ${Y}_{{\rm de},n+1}(s)$. 

\section{The Proof of Proposition~\ref{proof:align}}

\label{sec:appendixC}
\begin{proof}
       Enhance grid strength and system strength. Grid strength is given by:
\vspace{-1mm}
\begin{equation}\label{eq30:proof}
\begin{aligned}
\alpha(j\omega)&=\underline{\sigma}\left[{\bm B}^{N}_{\rm Grid}\otimes I_2-{\bm B}_{eq} \otimes {S}_v(j\omega)\right]\\
&=\underline{\sigma}\left[\tilde{\bm B}^{N}_{\rm Grid} \otimes I_2+ {\bm B}_{eq}\otimes \left(I_2-{S}_v(j\omega)\right)\right]\\
&\geq \underline{\lambda}\left[\tilde{\bm B}^{N}_{\rm Grid} \otimes I_2+ {\bm B}_{eq}\otimes \left(I_2-{S}_{v,H}(j\omega)\right)\right]\\
&\geq \underline{\sigma}\left(\tilde{\bm B}^{N}_{\rm Grid}\right)+{\rm const.}\times\left(1-\bar{\sigma}({S}_v(j\omega))\right)
\end{aligned}
\end{equation}
where, $\tilde{\bm B}^{N}_{\rm Grid}={\bm B}^{N}_{\rm Grid}-{\bm B}_{eq}$, ${S}_{v,H}=\frac{{S}_{v}+{S}^H_{v}}{2}$, and $\bar{\sigma}[{S}_v(j\omega)]=FI(j\omega)$. $\underline{\sigma}\left(\tilde{\bm B}^{N}_{\rm Grid}\right)$ is the grid strength prior to the $n+1$-th converter integration. If the eigenvectors of matrix $\tilde{\bm B}^{N}_{\rm Grid}$ are not equal to the eigenvectors corresponding to the zero eigenvalue of matrix ${\bm B}_{eq}$ (usually satisfied), $\rm const.$ is a constant greater than 0.

If $FI(j\omega)<1$, $\alpha(j\omega)>\underline{\sigma}\left(\tilde{\bm B}^{N}_{\rm Grid}\right)$. Reducing $FI(j\omega)$ increases grid strength $\alpha(j\omega)$, thereby improving the lower bound of system strength $\kappa(j\omega)$ by Proposition~\ref{grid strength}.
    
    Enhance bus strength. By bounding the singular values, we obtain:
\vspace{-1mm}
\begin{equation}\label{eq28:busproof}
\begin{aligned}
\underline{\sigma} \left[{Y}^{N+1}_{\rm Cl}(j\omega)\right]&\geq  \underline{\sigma}\left[{B}^{N+1}_{\rm Grid} \otimes \gamma(j\omega)\right] \overline{\sigma}^{-1}\left[{S}_v(j\omega)\right]-\overline{\sigma}\left[T_2(j\omega) \right]\\
\overline{\sigma}\left[{\bm Z}^{N,1}_{{\rm Cl},j}(j\omega)\right]&\leq \overline{\sigma}\left[A^{-1}_1(j\omega)A_2(j\omega)\right]\underline{\sigma} ^{-1}\left[{Y}^{N+1}_{\rm Cl}(j\omega)\right]
\,.
\end{aligned}
\end{equation}

Thus $\kappa_{n+1}(j\omega)$ is
\vspace{-1mm}
\begin{equation}\label{eq29:busin}
\begin{aligned}
\kappa_{n+1}(j\omega)&=\frac{1}{\textstyle \sum_{j=1}^{n}\overline{\sigma}\left[{\bm Z}^{N,1}_{{\rm Cl},j}(j\omega)\right]+\underline{\sigma}^{-1} \left[{Y}^{N+1}_{\rm Cl}(j\omega)\right]}\\
&\geq \frac{\underline{\sigma} \left[{Y}^{N+1}_{\rm Cl}(j\omega)\right]}{\textstyle \sum_{j=1}^{n}\overline{\sigma}_j\left[A^{-1}_1(j\omega)A_2(j\omega)\right]+1}\\
&\geq \frac{\underline{\sigma}\left[{B}^{N+1}_{\rm Grid} \otimes \gamma(j\omega)\right] \overline{\sigma}^{-1}\left[{S}_v(j\omega)\right]-\overline{\sigma}\left[T_2(j\omega) \right]}{\textstyle \sum_{j=1}^{n}\overline{\sigma}_j\left[A^{-1}_1(j\omega)A_2(j\omega)\right]+1}
\,,
\end{aligned}
\end{equation}
where $\overline{\sigma}_j(\cdot)$ denotes the maximum singular value of the $j$-th block matrix, and $\bar{\sigma}[{S}_v(j\omega)]=FI(j\omega)$. Reducing $FI(j\omega)$ increases the lower bound of $\underline{\sigma} \left[{Y}^{N+1}_{\rm Cl}(j\omega)\right]$ and the bus strength $\kappa_{n+1}(j\omega)$, and vice versa.

\end{proof}

\bibliographystyle{IEEEtran}
\vspace{-2mm}
\bibliography{IEEEabrv,RS}

@STRING{IEEE_J_ESTPE      = "{IEEE} Trans. Emerg. Sel. Topics Power Electron."}

@STRING{IEEE_J_EC         = "{IEEE} Trans. Energy Convers."}

@STRING{IEEE_J_PWRS       = "{IEEE} Trans. Power Syst."}

@STRING{IEEE_J_SG         = "{IEEE} Trans. Smart Grid"}

@INPROCEEDINGS{1:adaptive,
  author={Fradley, John and Preece, Robin and Barnes, Mike},
  booktitle={2019 IEEE Milan PowerTech}, 
  title={Adaptive Fast Frequency Response for Power Electronic Connected Energy Sources}, 
  year={2019},
  volume={},
  number={},
  pages={1-6},
  doi={10.1109/PTC.2019.8810597}}

@INPROCEEDINGS{20:MilanoPSCC,
  author={Milano, Federico and Dörfler, Florian and Hug, Gabriela and Hill, David J. and Verbič, Gregor},
  booktitle={2018 Power Systems Computation Conference (PSCC)}, 
  title={Foundations and Challenges of Low-Inertia Systems (Invited Paper)}, 
  year={2018},
  volume={},
  number={},
  pages={1-25},
  doi={10.23919/PSCC.2018.8450880}}

@ARTICLE{2,
  author={Bahrani, Behrooz and Ravanji, Mohammad Hasan and Kroposki, Benjamin and Ramasubramanian, Deepak and Guillaud, Xavier and Prevost, Thibault and Cutululis, Nicolaos-Antonio},
  journal={IEEE Power and Energy Magazine}, 
  title={Grid-Forming Inverter-Based Resource Research Landscape: Understanding the Key Assets for Renewable-Rich Power Systems}, 
  year={2024},
  volume={22},
  number={2},
  pages={18-29},
  doi={10.1109/MPE.2023.3343338}}

@ARTICLE{3:XIUQIANGDVOC,
  author={He, Xiuqiang and Huang, Linbin and Subotić, Irina and Häberle, Verena and Dörfler, Florian},
  journal={IEEE Trans. Power Electron.}, 
  title={Quantitative Stability Conditions for Grid-Forming Converters With Complex Droop Control}, 
  year={2024},
  volume={39},
  number={9},
  pages={10834-10852},
  doi={10.1109/TPEL.2024.3404251}}

@misc{GFM:NERC,
  title = {Grid forming technology},
  year = {2021},
  url ={https://www.nerc.com/Pages/default.aspx},
  note = {NERC report}
}

@misc{gfm:AEMO,
  title = {Voluntary Specification for Grid-forming Inverters},
  year = {2023},
  url ={https://aemo.com.au/-/media/files/initiatives/primary-frequency-response/2023/gfm-voluntary-spec.pdf},
  note = {AEMO report}
}

@misc{gfm:unifi,
  title = {Specifications for Grid-Forming Inverter-Based Resources },
  year = {2024},
  url ={https://docs.nrel.gov/docs/fy24osti/89269.pdf},
  note = {UNIFI report}
}

@misc{gfm:ACER,
  title = {ACER POLICY PAPER},
  year = {2022},
  url ={https://acer.europa.eu/sites/default/files/documents/Position%20Papers/260908%20ACER%20GCNCs%20Policy%20Paper_final.pdf},
  note = {ACER report}
}

@INPROCEEDINGS{6:Debryfrequencysmooth,
  author={Debry, Mane-Sophie and Denis, Guillaume and Prevost, Thibault},
  booktitle={2019 IEEE Milan PowerTech}, 
  title={Characterization of the Grid-Forming Function of a Power Source Based on its External Frequency Smoothing Capability}, 
  year={2019},
  volume={},
  number={},
  pages={1-6},
  doi={10.1109/PTC.2019.8810409}}

@article{21:grossfrequencysmooth,
  title={Input-Output Specifications of Grid-Forming Functions and Data-Driven Verification Methods},
  author={Bui, Jennifer T and Gro{\ss}, Dominic},
  journal={arXiv preprint arXiv:2404.15951},
  year={2024}
}

@ARTICLE{7:howmany,
  author={Xin, Huanhai and Liu, Chenxi and Chen, Xia and Wang, Yuxuan and Prieto-Araujo, Eduardo and Huang, Linbin},
  journal = IEEE_J_PWRS,
  title={How Many Grid-Forming Converters Do We Need? a Perspective From Small Signal Stability and Power Grid Strength}, 
  year={2024},
  volume={},
  number={},
  pages={1-13},
  doi={10.1109/TPWRS.2024.3393877}}

@article{8:eduardovoltage,
  title={A Dynamic Similarity Index for Assessing Voltage Source Behaviour in Power Systems},
  author={Alican, Onur and Moutevelis, Dionysios and Arevalo-Soler, Josep and Collados-Rodriguez, Carlos and Amoros-Torrent, Jaume and Gomis-Bellmunt, Oriol and Cheah-Mane, Marc and Prieto-Araujo, Eduardo},
  journal={arXiv preprint arXiv:2501.16167},
  year={2025}
}

@ARTICLE{9:pll-gfm,
  author={Schweizer, Mario and Almér, Stefan and Pettersson, Sami and Merkert, Arvid and Bergemann, Vivien and Harnefors, Lennart},
  journal={IEEE Trans. Power Electron.}, 
  title={Grid-Forming Vector Current Control}, 
  year={2022},
  volume={37},
  number={11},
  pages={13091-13106},
  doi={10.1109/TPEL.2022.3177938}}

@misc{10,
  title = {Dynamic assessment of Power System Strength in systems with a large share of generation from renewable sources},
  year = {2024},
  url = {https://www.e-cigre.org/publications.html},
  note = {CIGRE report}
}

@misc{10:cigre,
  title = {Dynamic assessment of Power System Strength in systems with a large share of generation from renewable sources},
  year = {2024},
  url = {https://www.e-cigre.org/publications.html},
  note = {CIGRE report}
}

@misc{strength:aemo,
  title = {System Strength Impact Assessment Guidelines},
  year = {2024},
  url = {https://aemo.com.au/energy-systems/electricity/national-electricity-market-nem/participate-in-the-market/network-connections/system-strength-impact-assessment-guidelines},
  note = {AEMO report}
}

@ARTICLE{xiongfei:passive,
  author={Chen, Feifan and Khong, Sei Zhen and Harnefors, Lennart and Wang, Xiongfei and Wang, Dan and Sandberg, Henrik and Zhao, Liang and Routimo, Mikko and Kukkola, Jarno and Sou, Kin Cheong and Johansson, Karl Henrik},
  journal={IEEE Trans. Power Electron.}, 
  title={An Extended Frequency-Domain Passivity Theory for MIMO Dynamics Specifications of Voltage-Source Inverters}, 
  year={2025},
  volume={40},
  number={2},
  pages={2943-2957},
  doi={10.1109/TPEL.2024.3488853}}

@ARTICLE{guyunjie:impedance,
  author={Gu, Yunjie and Li, Yitong and Zhu, Yue and Green, Timothy C.},
  journal={IEEE Transactions on Power Systems}, 
  title={Impedance-Based Whole-System Modeling for a Composite Grid via Embedding of Frame Dynamics}, 
  year={2021},
  volume={36},
  number={1},
  pages={336-345},
  doi={10.1109/TPWRS.2020.3004377}}

@ARTICLE{11,
  author={Yang, Chaoran and Huang, Linbin and Xin, Huanhai and Ju, Ping},
  journal={IEEE Trans. Power Syst.}, 
  title={Placing Grid-Forming Converters to Enhance Small Signal Stability of PLL-Integrated Power Systems}, 
  year={2021},
  volume={36},
  number={4},
  pages={3563-3573},
  doi={10.1109/TPWRS.2020.3042741}}

@ARTICLE{12,
  author={Zhao, Fangzhou and Zhu, Tianhua and Harnefors, Lennart and Fan, Bo and Wu, Heng and Zhou, Zichao and Sun, Yin and Wang, Xiongfei},
  journal={IEEE Open Journal of Power Electronics}, 
  title={Closed-Form Solutions for Grid-Forming Converters: A Design-Oriented Study}, 
  year={2024},
  volume={5},
  number={},
  pages={186-200},
  doi={10.1109/OJPEL.2024.3357128}}

@ARTICLE{14,
  author={Me, Si Phu and Ravanji, Mohammad Hasan and Leonardi, Bruno and Ramasubramanian, Deepak and Ma, Jin and Zabihi, Sasan and Bahrani, Behrooz},
  journal={IEEE Transactions on Power Electronics}, 
  title={Transient Stability Analysis of Virtual Synchronous Generator Equipped With Quadrature-Prioritized Current Limiter}, 
  year={2023},
  volume={38},
  number={9},
  pages={10547-10553},
  doi={10.1109/TPEL.2023.3278835}}

@ARTICLE{xiongfei:syn_overview,
  author={Wang, Xiongfei and Taul, Mads Graungaard and Wu, Heng and Liao, Yicheng and Blaabjerg, Frede and Harnefors, Lennart},
  journal={IEEE Open Journal of Industry Applications}, 
  title={Grid-Synchronization Stability of Converter-Based Resources—An Overview}, 
  year={2020},
  volume={1},
  number={},
  pages={115-134},
  doi={10.1109/OJIA.2020.3020392}}

@ARTICLE{kunder:stability_classification,
  author={Kundur, P. and Paserba, J. and Ajjarapu, V. and Andersson, G. and Bose, A. and Canizares, C. and Hatziargyriou, N. and Hill, D. and Stankovic, A. and Taylor, C. and Van Cutsem, T. and Vittal, V.},
  journal=IEEE_J_PWRS, 
  title={Definition and classification of power system stability IEEE/CIGRE joint task force on stability terms and definitions}, 
  year={2004},
  volume={19},
  number={3},
  pages={1387-1401},
  doi={10.1109/TPWRS.2004.825981}}

@ARTICLE{Linbin:PLL_sy,
  author={Huang, Linbin and Xin, Huanhai and Li, Zhiyi and Ju, Ping and Yuan, Hui and Lan, Zhou and Wang, Zhen},
  journal=IEEE_J_SG, 
  title={Grid-Synchronization Stability Analysis and Loop Shaping for PLL-Based Power Converters With Different Reactive Power Control}, 
  year={2020},
  volume={11},
  number={1},
  pages={501-516},
  doi={10.1109/TSG.2019.2924295}}

@INPROCEEDINGS{Thomas:GFM_GFL_comparison,
  author={Pattabiraman, Dinesh and Lasseter., R. H. and Jahns, T. M.},
  booktitle=IEEE_PESGM, 
  title={Comparison of Grid Following and Grid Forming Control for a High Inverter Penetration Power System}, 
  year={2018},
  volume={},
  number={},
  pages={1-5},
  doi={10.1109/PESGM.2018.8586162}}

@ARTICLE{Huanhai:gSCR,
  author={Dong, Wei and Xin, Huanhai and Wu, Di and Huang, Linbin},
  journal=IEEE_J_PWRS, 
  title={Small Signal Stability Analysis of Multi-Infeed Power Electronic Systems Based on Grid Strength Assessment}, 
  year={2019},
  volume={34},
  number={2},
  pages={1393-1403},
  doi={10.1109/TPWRS.2018.2875305}}

@ARTICLE{WenBo:dq_model,
  author={Wen, Bo and Burgos, Rolando and Boroyevich, Dushan and Mattavelli, Paolo and Shen, Zhiyu},
  journal=IEEE_J_ESTPE, 
  title={AC Stability Analysis and dq Frame Impedance Specifications in Power-Electronics-Based Distributed Power Systems}, 
  year={2017},
  volume={5},
  number={4},
  pages={1455-1465},
  doi={10.1109/JESTPE.2017.2728640}}

@ARTICLE{Huanhai:place_GFM,
  author={Yang, Chaoran and Huang, Linbin and Xin, Huanhai and Ju, Ping},
  journal=IEEE_J_PWRS, 
  title={Placing Grid-Forming Converters to Enhance Small Signal Stability of PLL-Integrated Power Systems}, 
  year={2021},
  volume={36},
  number={4},
  pages={3563-3573},
  doi={10.1109/TPWRS.2020.3042741}}

@ARTICLE{kehao:dual_axis,
  author={Xin, Huanhai and Zhuang, Kehao and Hu, Pengfei and Huang, Linbin and Liu, Xinyu and Hu, Guang},
  journal=IEEE_J_EC, 
  title={Small Signal Synchronization Stability Analysis of Interconnected System Containing Grid-Forming and Grid-Following Converters from the Perspective of Dual-axis Synchronous Generator}, 
  year={2024},
  volume={},
  number={},
  pages={1-14},
  doi={10.1109/TEC.2024.3487201}}

@INPROCEEDINGS{1,
  author={Fradley, John and Preece, Robin and Barnes, Mike},
  booktitle={2019 IEEE Milan PowerTech}, 
  title={Adaptive Fast Frequency Response for Power Electronic Connected Energy Sources}, 
  year={2019},
  volume={},
  number={},
  pages={1-6},
  doi={10.1109/PTC.2019.8810597}}

@INPROCEEDINGS{6,
  author={Debry, Mane-Sophie and Denis, Guillaume and Prevost, Thibault},
  booktitle={2019 IEEE Milan PowerTech}, 
  title={Characterization of the Grid-Forming Function of a Power Source Based on its External Frequency Smoothing Capability}, 
  year={2019},
  volume={},
  number={},
  pages={1-6},
  doi={10.1109/PTC.2019.8810409}}

@article{21,
  title={Input-Output Specifications of Grid-Forming Functions and Data-Driven Verification Methods},
  author={Bui, Jennifer T and Gro{\ss}, Dominic},
  journal={arXiv preprint arXiv:2404.15951},
  year={2024}
}

@ARTICLE{GFM:keytechonology,
  author={Matevosyan, Julia and Badrzadeh, Babak and Prevost, Thibault and Quitmann, Eckard and Ramasubramanian, Deepak and Urdal, Helge and Achilles, Sebastian and MacDowell, Jason and Hsien Huang, Shun and Vital, Vijay and O’Sullivan, Jon and Quint, Ryan},
  journal={IEEE Power and Energy Magazine}, 
  title={Grid-Forming Inverters: Are They the Key for High Renewable Penetration?}, 
  year={2023},
  volume={21},
  number={2},
  pages={77-86},
  doi={10.1109/MPAE.2023.10083082}}

@book{skogestad2005multivariable,
  title     = {Multivariable Feedback Control: Analysis and Design},
  author    = {Skogestad, Sigurd and Postlethwaite, Ian},
  year      = {2005},
  publisher = {John Wiley \& Sons},
  edition   = {2nd}
}

@ARTICLE{gainandphase,
  author={Huang, Linbin and Wang, Dan and Wang, Xiongfei and Xin, Huanhai and Ju, Ping and Johansson, Karl H. and Dörfler, Florian},
  journal={IEEE Trans. Power Syst.},
  title={Gain and Phase: Decentralized Stability Conditions for Power Electronics-Dominated Power Systems}, 
  year={2024},
  volume={39},
  number={6},
  pages={7240-7256},
  doi={10.1109/TPWRS.2024.3380528}}

@ARTICLE{crossforming,
  author={He, Xiuqiang and Desai, Maitraya Avadhut and Huang, Linbin and Dörfler, Florian},
  journal={IEEE Trans. Power Electron.}, 
  title={Cross-Forming Control and Fault Current Limiting for Grid-Forming Inverters}, 
  year={2025\color{black}},
  volume={40},
  number={3},
  pages={3980-4007},
  doi={10.1109/TPEL.2024.3500885}}

@misc{passiveofgfm,
      title={Dynamic Passivity Multipliers for Plug-and-Play Stability Certificates of Converter-Dominated Grids}, 
      author={Andrey Gorbunov and Youhong Chen and Petr Vorobev and Jin Ma and Gregor Verbic},
      year={2026\color{black}},
      eprint={2602.09150},
      archivePrefix={arXiv},
      primaryClass={eess.SY},
      url={https://arxiv.org/abs/2602.09150}, 
}

@ARTICLE{PLLGFM,
   author={ Harnefors, Lennart and Schweizer, Mario and Kukkola, Jarno and Routimo, Mikko and Hinkkanen, Marko and Wang, Xiongfei},
  journal={IEEE Trans. Power Electron.},
  title={Generic PLL-Based Grid-Forming Control}, 
  year={2022\color{black}},
  volume={37},
  number={2},
  pages={1201-1204},
  doi={10.1109/TPEL.2021.3106045}}

@ARTICLE{strength_magazine,
  author={Huang, Linbin and He, Xiuqiang and Xin, Huanhai and Li, Zhiyi and Ju, Ping and Ma, Fuyilong and Wang, Kang and Dörfler, Florian},
  journal={IEEE Power and Energy Magazine}, 
  title={System Strength in Power Electronics-Dominated Power Systems: An Enabler to Stability and Control}, 
  year={2026\color{black}},
  volume={24},
  number={1},
  pages={49-66},
  doi={10.1109/MPE.2025.3599131}}

@INPROCEEDINGS{sgimpedance,
  author={Aalam, Mohammad Nair and Tang, Ye and Thacker, Timothy and Burgos, Rolando},
  booktitle={2025 IEEE Energy Conversion Conference Congress and Exposition (ECCE)}, 
  title={dq-Frame Small-Signal Impedance Modeling of a Synchronous Machine}, 
  year={2025},
  volume={},
  number={},
  pages={1-7},
  doi={10.1109/ECCE58356.2025.11259873}}

@ARTICLE{dvoc_dq,
  author={Zeng, Zheran and Fan, Jiayu and Sun, Yin and Wang, Songda and Yang, Dongsheng},
  journal={IEEE Trans. Power Electron.}, 
  title={A Dispatchable Virtual Oscillator Controller in the dq Frame With Enhanced Grid-Forming Power Reference Tracking Capability}, 
  year={2025},
  volume={40},
  number={8},
  pages={10973-10987},
  doi={10.1109/TPEL.2025.3556325}}

@ARTICLE{SG_model,
  author={Report, IEEE Committee},
  journal={IEEE Trans. Power App. Syst.}, 
  title={Dynamic Models for Steam and Hydro Turbines in Power System Studies}, 
  year={1973},
  volume={PAS-92},
  number={6},
  pages={1904-1915},
  doi={10.1109/TPAS.1973.293570}}

\begin{IEEEbiography}[{\includegraphics[width=1in,height=1.25in,clip,keepaspectratio]{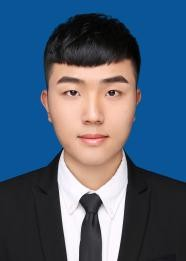}}]{Kehao Zhuang}
(Member, IEEE) was born in Shandong Province, China, in 1998. He received his B.S. and Ph.D. degrees from Northeast Electric Power University and Zhejiang University, China, in 2020 and 2025, respectively. He is currently a Postdoctoral Researcher at ETH Zürich with the Automatic Control Laboratory. His research interests include the stability analysis and control of power electronics-dominated power system.\end{IEEEbiography}

\begin{IEEEbiography}[{\includegraphics[width=1in,height=1.25in,clip,keepaspectratio]{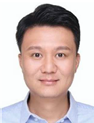}}]{Huanhai Xin}
(Senior Member, IEEE) received the Ph.D. degree in electrical engineering from the College of Electrical Engineering, Zhejiang University, Hangzhou, China, in 2007. He is currently a Professor with the Department of Electrical Engineering, Zhejiang University. From 2009 to 2010, he was a Postdoctoral Researcher with the Department of Electrical Engineering and Computer Science, University of Central Florida, Orlando, FL, USA. His re search interests include ac–dc power system transient stability analysis and control, and grid-integration of large-scale renewable energy to weak grids.\end{IEEEbiography}

\begin{IEEEbiography}[{\includegraphics[width=1in,height=1.25in,clip,keepaspectratio]{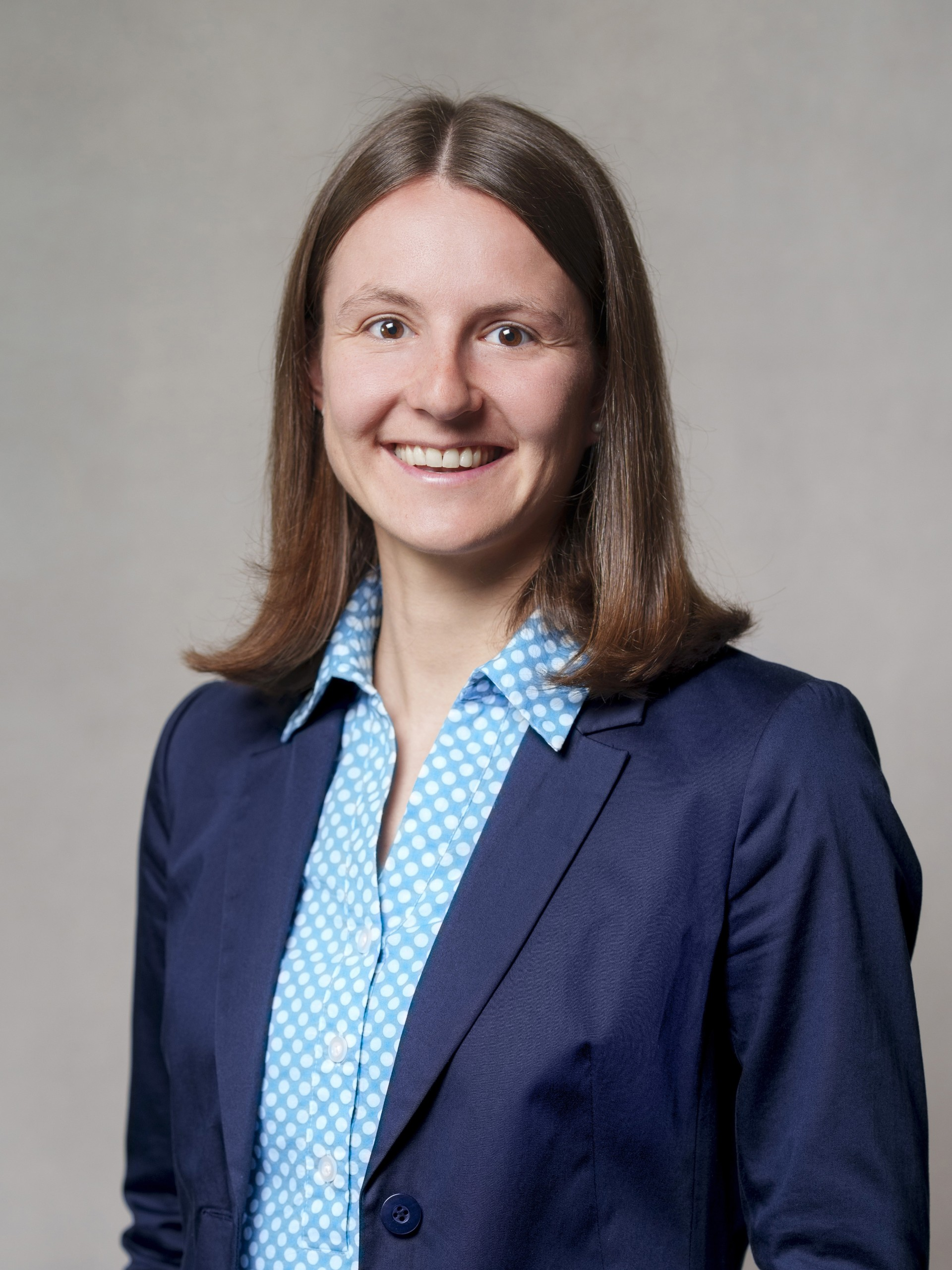}}]{Verena Häberle}
(Member, IEEE) received the B.Sc., M.Sc., and Ph.D. degrees from ETH Zürich, Switzerland, in 2018, 2020, and 2025, respectively. She is currently a Postdoctoral Researcher at ETH Zürich with the Power Systems Laboratory and the Automatic Control Laboratory. In fall 2024, she was a visiting researcher at the California Institute of Technology. For her outstanding academic achievements during her Ph.D. under Professor Florian Dörfler, she was honoured with the Hans Eggenberger Award. Her research focuses on dynamic ancillary services, virtual power plants, and the control and interoperability certification of power electronics–dominated power systems.  \end{IEEEbiography}

\begin{IEEEbiography}[{\includegraphics[width=1in,height=1.25in,clip,keepaspectratio]{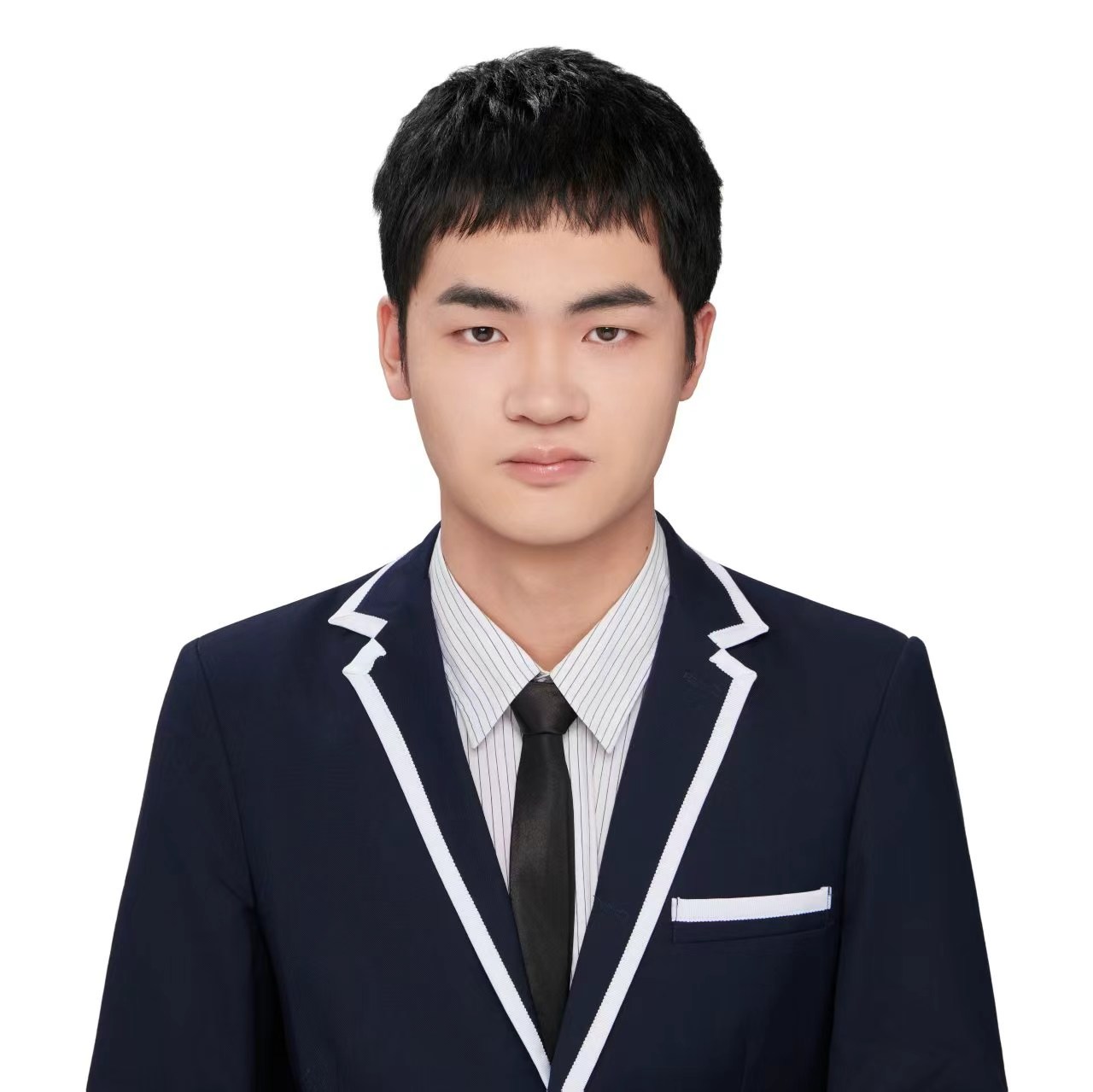}}]{Liangxiao Luo}
received the B.Eng. degree in electrical engineering from Sichuan University, Chengdu, China, in 2024. He is currently working toward the Ph.D. degree with the College of Electrical Engineering, Zhejiang University, Hangzhou, China. His research interests include stability analysis and control of renewable power systems. \end{IEEEbiography}

\begin{IEEEbiography}[{\includegraphics[width=1in,height=1.25in,clip,keepaspectratio]{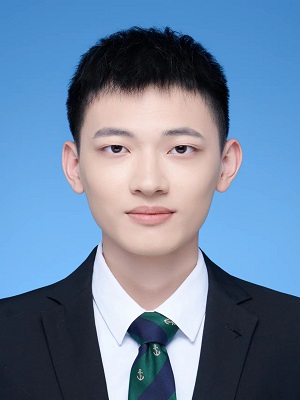}}]{Xiaoying Liu}
received the B.S. degree in power system and automation from North China Electric Power University (NCEPU), China, in 2021. He is currently pursuing the Ph.D. degree with the College of Electrical Engineering, Zhejiang University, Hangzhou, China. His research interests include fundamental theories for stability analysis and control of power-electronics-dominated renewable energy power systems.\end{IEEEbiography}

\begin{IEEEbiography}[{\includegraphics[width=1in,height=1.25in,clip,keepaspectratio]{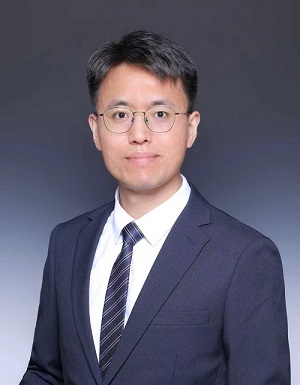}}]{Xiuqiang He}
(Member, IEEE) received the B.S. and Ph.D. degrees in Control Science and Engineering from Tsinghua University, Beijing, China, in 2016 and 2021, respectively. Since 2021, he has been a Postdoctoral Researcher and then was promoted to a Senior Scientist with the Automatic Control Laboratory, ETH Zürich, Switzerland. He is currently an Assistant Professor with the Department of Automation, Tsinghua University, China. His research interests include power system dynamics, stability, control, and lunar surface microgrid control. Dr. He was the recipient of Beijing Outstanding Graduates Award and the Outstanding Doctoral Dissertation Award from Tsinghua University.\end{IEEEbiography}

\begin{IEEEbiography}[{\includegraphics[width=1in,height=1.25in,clip,keepaspectratio]{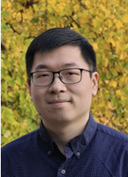}}]{Linbin Huang}
(Member, IEEE) received the B.Eng. and Ph.D. degrees from Zhejiang University, Hangzhou, China, in 2015 and 2020, respectively. From 2020 to 2024, he was a senior scientist with the Automatic Control Laboratory at ETH Zurich, 8092 Zurich, Switzerland. Currently, he is a Professor with the College of Electrical Engineering, Zhejiang University. His research interests include power system stability, optimal control of power electron ics, and data-driven control.\end{IEEEbiography}

\begin{IEEEbiography}[{\includegraphics[width=1in,height=1.25in,clip,keepaspectratio]{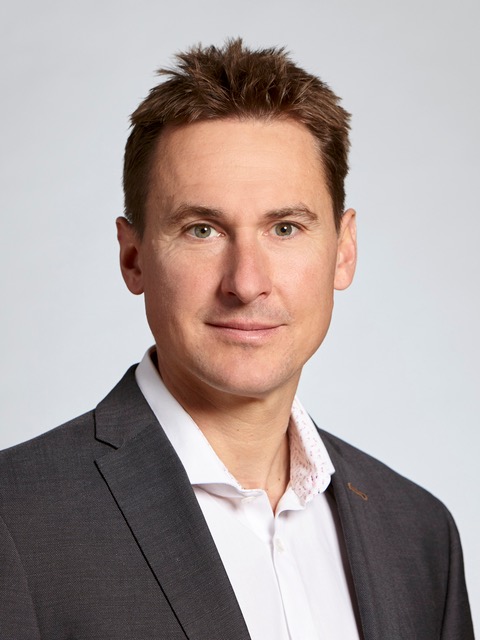}}]{Florian Dörfler}
(Fellow, IEEE) is a Professor at the Automatic Control Laboratory at ETH Zürich. He received his Ph.D. degree in Mechanical Engineering from the University of California at Santa Barbara in 2013, and a Diplom degree in Engineering Cybernetics from the University of Stuttgart in 2008. From 2013 to 2014 he was an Assistant Professor at the University of California Los Angeles. He is a recipient of the 2025 Rössler Prize, the highest scientific award at ETH Zürich across all disciplines, as well as the distinguished career awards by IFAC (Manfred Thoma Medal 2020) and EUCA (European Control Award 2020). He and his team were recipients of the 2011 O. Hugo Schuck Best Paper Award, the 2012-2014 Automatica Best Paper Award, the 2016 IEEE Circuits and Systems Guillemin-Cauer Best Paper Award, the 2022 IEEE Transactions on Power Electronics Prize Paper Award, the 2024 Control Systems Magazine Outstanding Paper Award, multiple Best PhD thesis awards at UC Santa Barbara and ETH Zürich, winners or finalists for Best Student Paper awards at the European Control Conference (2013, 2019), the American Control Conference (2010, 2016, 2024), the Conference on Decision and Control (2020), the PES General Meeting (2020), the PES PowerTech Conference (2017, 2025), the International Conference on Intelligent Transportation Systems (2021), the IEEE CSS Swiss Chapter Young Author Best Journal Paper Award (2022, 2024, 2025), the IFAC Conferences on Nonlinear Model Predictive Control (2024) and Cyber-Physical-Human Systems (2024), and NeurIPS Oral (2024). He is a Fellow of the IEEE and currently serving on the council of the European Control Association and as a senior editor of Automatica. \end{IEEEbiography}

\end{document}